\documentstyle[12pt,aasms]{article}

\begin{document}

\title{Nucleosynthesis in Neutrino-Driven Winds: I. The Physical Conditions}

\author{Y.-Z. Qian}
\affil{Physics Department, 161-33, California Institute of Technology,
Pasadena, CA 91125}

\and

\author{S. E. Woosley}

\affil{Board of Studies in Astronomy and Astrophysics, University of
California, Santa Cruz,\\Santa Cruz, CA 95064\\and\\
Max Planck Institut f\"ur Astrophysik, 85740 Garching bei M\"unchen,
Germany}

\begin{abstract} 

During the first 20 seconds of its life, the enormous neutrino luminosity
of a neutron star drives appreciable mass loss from its surface. Previous
investigations have shown that this neutrino-driven wind could be the site
where the $r$-process occurs. The nucleosynthesis is sensitive to four
physical parameters characterizing the wind: its mass loss rate, the
entropy per baryon, the electron fraction, and the dynamic time
scale. Different authors, using numerical models for supernovae, have
arrived at qualitatively different values for these key parameters. Here we
derive their values analytically and test our analytic results by numerical
calculations using an implicit hydrodynamic code.  Employing our analytic
and numerical methods, we also investigate how various factors can affect
our results. The derived entropy typically falls short, by a factor of two
to three, of the value required to produce a strong $r$-process. Various
factors that might give a higher entropy or a more rapid expansion
in the wind are
discussed.

\end{abstract}


\section{Introduction}

Recent years have seen progress in our understanding of both the Type II
supernova mechanism and heavy element nucleosynthesis. An intriguing
suggestion has been that the origin of the $r$-process is in the
neutrino-driven wind of a young neutron star (Woosley \& Hoffman 1992, see
also Meyer et al. 1992; Howard et al. 1993; Witti, Janka, \& Takahashi 1994;
and Takahashi, Witti, \& Janka 1994). The winds leave the neutron star
about 1 to 20 s after its creation by the stellar core collapse. The mass
loss is sustained by neutrino heating during the Kelvin-Helmholtz cooling
phase of the neutron star. Recently, two groups (Woosley et al. 1994;
Takahashi et al. 1994) have published calculations of $r$-process in these
neutrino-heated ejecta based upon different supernova models. These
calculations yield tantalizing results compared with the observed solar
system abundance distribution of $r$-process elements, but the calculations
share some drawbacks and exhibit differences.

One failure common to all calculations so far is the overproduction of
nuclei in the vicinity of the magic neutron shell, $N=50$, specifically
$^{88}$Sr, $^{89}$Y, and $^{90}$Zr. Interestingly, this difficulty can be
circumvented and even used to provide some attractive nucleosynthesis of the
light $p$-process nuclei if the electron fraction $Y_e$ in the ejecta is
increased slightly from the value found by Wilson in Woosley et al. (1994).
Hoffman et al. (1996a) have shown that for $0.484\lesssim Y_e\lesssim 0.488$,
the problematic nucleosynthesis disappears. This result highlights the need
for a detailed and accurate analysis of $Y_e$ obtained in the wind, and also
hints at some critical uncertainties in the supernova models.

A second difficulty is that various calculations do not agree on the entropy
in the wind at the time the $r$-process is alleged to occur. Woosley et al.
(1994)
based their calculations on a supernova model computed by Jim Wilson, where
high entropies, $\sim 400\ k$ per baryon (hereafter, we frequently refer to
entropy without its units in Boltzmann constant $k$ per baryon), 
were obtained at times later
than about 10 seconds. These high entropies were very beneficial, perhaps
even necessary for the production of nuclei with mass number $A \gtrsim 130$ and
especially the $r$-process abundance peak at $A\sim 195$.  However, in the
supernova model evolved by Janka and employed in the calculations of Witti et
al. (1994) and Takahashi et al. (1994), entropies of only $\lesssim 100$ 
were obtained. In
fact, Takahashi et al. (1994) had to artificially impose an increase 
by a factor of
$\sim 5$ in the entropy to obtain a successful $r$-process. Similar small
values of the entropy were also obtained in numerical calculations by Woosley
(1993 and unpublished). This difference further demonstrates the necessity of
an unambiguous understanding of the physical conditions obtained in the
neutrino-driven wind.

Finally, Meyer (1995) has recently shown that the $r$-process in the
neutrino-heated supernova ejecta might be seriously affected by neutrino
spallation reactions on $\alpha$-particles under conditions obtained in
Wilson's supernova model. These reactions were neglected in the previous
$r$-process calculations. Inclusion of these reactions reduces the 
production of nuclei with
$A\sim 195$ in Wilson's supernova model. To counter the damaging effects of
neutrino spallation reactions on $\alpha$-particles requires an entropy even
higher than $\sim 400$, a lower electron fraction, a shorter dynamic time
scale, a larger distance of the $r$-process site from the neutrino source, or
some combination of the above.

It is clear then that a careful analysis of the physical conditions in the
neutrino-driven wind is needed. These conditions include the
entropy per baryon, the electron fraction, the dynamic time scale, and the
mass outflow rate. While simple analytic calculations cannot give the full
detail of a multidimensional numerical study with non-LTE treatments of
neutrino transport, they can be very useful for understanding the output
from the numerical codes and delimiting, from first principles, 
likely ranges for
the conditions. That is the chief goal of this paper.

Because the conditions relevant for heavy element nucleosynthesis exist, in
terms of neutron star time scales, long after the shock has been launched in
a successful supernova explosion, the neutrino-driven wind can be reasonably
approximated as a quasi-steady spherical outflow. This assumption is borne
out by the recent two-dimensional numerical calculations by Janka and
M\"{u}ller (1995). This fact facilitates analytic studies of the
neutrino-driven wind, and greatly simplifies the numerical treatment (though
we shall also briefly examine non-steady outflow). It also has the
consequence that our key results are independent of the details of how the
shock gets launched.

In $\S2$, we set up the basic equations for analytic studies of the
neutrino-driven wind. Our work here is similar in ways to that
of Duncan, Shapiro, and Wasserman (1986), but more extensive and directed
towards nucleosynthetic issues. The relevant initial and boundary
conditions and the input neutrino physics to solve these equations are
discussed, and we present the analytic results for the entropy per baryon,
the dynamic time scale, and the mass outflow rate in the ejecta in
$\S3$. In $\S4$, we compare these analytic results with numerical
calculations using an implicit hydrodynamic code KEPLER while varying the
neutron star mass, radius, and neutrino luminosity. Some dependencies upon
the Newtonian approximation, the steady-state assumption, boundary
pressure, and additional energy sources are also briefly explored. In
$\S5$, we discuss how the electron fraction $Y_e$ is determined in the
ejecta, and present a novel neutrino ``two-color plot" that aids in
understanding, qualitatively, the nature of heavy element nucleosynthesis
in supernovae. In $\S6$, we discuss the dependence of a successful
$r$-process on the set of three physical parameters: the entropy per
baryon, the electron fraction, and the dynamic time scale, and speculate
further on how various factors can affect the properties of the
neutrino-driven wind. In a separate paper (Hoffman et
al. 1996b), we will give a detailed discussion of the implications of our
results for heavy element nucleosynthesis in supernovae.

\section{Equations to model the neutrino-driven wind}

The time of interest here is later than about 1 s after the shock has been
launched in a successful Type II or Ib supernova. By this time the shock
wave responsible for the supernova explosion has reached a radius of about
10 thousand kilometers and the temperature at that shock has declined to
about a billion degrees. A hot neutron star with a radius of $\sim 10$ km
lies near the center of this supernova. Its Kelvin-Helmholtz phase of
cooling by neutrino emission lasts about 10 s, during which time the
gravitational binding energy of the final neutron star, a few times $10^{53}$
erg, is radiated approximately equally in $\nu_e$, $\bar\nu_e$, $\nu_\mu$,
$\bar\nu_\mu$, $\nu_\tau$, and $\bar\nu_\tau$ (Janka 1995).
Therefore, the average luminosity for a specific neutrino species at this
late time is of order 10$^{51}$--$10^{52}$ erg s$^{-1}$. As this neutrino
flux streams through the region above the neutron star, the following weak
interactions can take place: absorption of $\nu_e$ and $\bar\nu_e$,
neutrino-electron scattering, and neutrino-antineutrino annihilation.  The
last two are experienced by all flavors of neutrinos. These interactions
transfer energy from neutrinos to the matter above the neutron star and heat
it. Consequently, this material expands away from the neutron star and
develops into a mass outflow we call the ``neutrino-driven wind."

Intuitively, once the convection associated with the launching of the shock
has ceased, one expects this mass outflow to be relatively smooth, and
adequately described by a quasi-steady-state approximation. In fact, this
expectation is verified by recent two-dimensional numerical simulations
(Janka and M\"{u}ller 1995). This is because during the Kelvin-Helmholtz
cooling phase of the neutron star, (i) the global characteristics of the
neutrino flux, e.g., luminosity and energy distribution, change rather
slowly; (ii) the properties of the neutron star, e.g., mass, radius, and
surface temperature, also evolve very slowly; and (iii) the supernova shock
wave is at very large radii, and has little influence on the conditions
close to the neutron star.

Such is the ideal world.  In fact, the real local situation in the wind may
be more complex. Woosley \& Weaver (1995) find considerable mass continues
to accrete onto the neutron star for several hours following its formation.
Convection may also continue at some level even 10 s after the shock is
launched. Rotation may lead to a breaking of spherical symmetry. However,
in order to proceed, we shall adopt the steady-state assumption, in which
case the dynamic equations can be written as (Duncan et al.
1986): 
\begin{eqnarray} 
\dot M &=& 4\pi r^2\rho v,\\ v{dv\over dr} &=&
-{1\over\rho}{dP\over dr}-{GM\over r^2},\\ \dot q &=&
v\left({d\epsilon\over dr}-{P\over\rho^2}{d\rho\over dr}\right),
\end{eqnarray} 
where $\rho$ is the rest mass density, $v$ is the outflow
velocity, $T$ is the temperature, $P$ and $\epsilon$ are the total pressure
and specific internal energy corresponding to non-relativistic matter and
relativistic particles in the outflow, respectively, $\dot q$ is the net
specific heating rate due to neutrino interactions above the neutron star,
$M$ is the mass of the neutron star, and $\dot M$ is the
constant mass outflow rate in the ejecta. In writing the above equations,
we have made the following assumptions: (i) the outflow velocity is
non-relativistic; (ii) the general relativistic effects above the neutron
star are small; and (iii) the neutron star is the predominant source of
gravity. We will discuss the validity of these assumptions later.

As we will see in $\S3$ and $\S4$, the characteristics of the neutrino-driven
wind are mostly determined at $T\gtrsim 0.5$ MeV. At these temperatures, the
material is composed of non-relativistic free neutrons and protons,
relativistic electrons and positrons, and photon radiation. The expression
for pressure, $P$,  and internal energy, $\epsilon$, can then be
approximately written as
\begin{eqnarray}
P &=& {11\pi^2\over 180}T^4\left(1+{30\eta^2\over 11\pi^2}+
{15\eta^4\over 11\pi^4}\right)
+{\rho\over m_N} T,\\
\epsilon &=& {11\pi^2\over 60}{T^4\over\rho}\left(1+{30\eta^2\over 11\pi^2}
+{15\eta^4\over 11\pi^4}\right)+{3\over 2}{T\over m_N}, 
\end{eqnarray}
where $m_N$ is the nucleon rest mass, and $\eta=\mu/T$ is the degeneracy 
parameter with $\mu$ the electron chemical
potential. The degeneracy parameter, $\eta$, is related to the electron
fraction $Y_e$ through
\begin{equation}
Y_e={T^3\over (\rho/m_N)}{\eta\over 3}\left(1+{\eta^2\over\pi^2}\right).
\end{equation}
In equations (4) to (6), we have used the units in which the Planck constant
$\hbar$,
the speed of light $c$, and the Boltzmann constant $k$ are taken to be unity.
It is easy to obtain the correct dimension of a physical quantity if
one knows that $\hbar c=197.33$ MeV fm. For example, equation (6)
in normal units would be $Y_e=(kT/\hbar c)^3(m_N/\rho)(\eta/3)(1+\eta^2/\pi^2)$.

The electron fraction $Y_e$ is, in turn, determined by
\begin{equation}
v{dY_e\over dr}=\lambda_{\nu_en}+\lambda_{e^+n}-(\lambda_{\nu_en}+
\lambda_{e^+n}+\lambda_{\bar\nu_ep}+\lambda_{e^-p})Y_e,
\end{equation}
where $\lambda_{\nu_en}$, $\lambda_{\bar\nu_ep}$, $\lambda_{e^+n}$, and
$\lambda_{e^-p}$ are the rates for the forward and reverse reactions in the
following equations:
\begin{mathletters}
\begin{eqnarray}
\nu_e+n &\rightleftharpoons& p+e^-,\\
\bar\nu_e+p &\rightleftharpoons& n+e^+.
\end{eqnarray}
\end{mathletters}

Given the input neutrino physics (i.e., the expressions of $\dot q$, 
$\lambda_{\nu_en}$,
$\lambda_{\bar\nu_ep}$, $\lambda_{e^+n}$, and $\lambda_{e^-p}$), the initial 
conditions at the neutron star surface,
and the boundary conditions at the shock wave, we can think of equations (1)
through (7) as a complete set of equations for an ``eigenvalue'' problem
of $\dot M$ in the neutrino-driven wind. Sometimes it is
convenient to think of these equations as describing a Lagrangian mass
element moving away from the neutron star with velocity $v(r)$. In this case,
we can introduce a time variable $t=\int^r_{r_0}dr/v(r)$, with $r_0$ being
some initial reference radius. Then the derivative with respect to $t$ is
equivalent to $v(d/dr)$. 

\section{Analytic description of the neutrino-driven wind}
It is helpful to estimate the general conditions before a detailed description
of the neutrino-driven wind. The
ejecta leave the neutron star surface with a small, very subsonic initial
velocity. In order to escape to large radii, a nucleon has to gain enough
energy from the neutrino flux to overcome its gravitational potential at
the neutron star surface. For a typical neutron star, the mass is
$M\sim 1.4\ M_{\sun}$, and the radius is $R\sim 10$ km. The amount of
energy provided by neutrino heating for a nucleon has to be at least $\sim
GMm_N/R\sim 200$ MeV. Because the neutrino flux decreases as $r^{-2}$ away
from the neutron star, we expect that most of the heating takes place close
to the neutron star. The surface temperature of a nascent neutron star is
several MeV, and the thermal kinetic energy of a nucleon close to the
neutron star is of the same order. Since the initial velocity of the nucleon
is also small, the nucleon is incapable of carrying the amount of energy
obtained from the neutrino flux.  Almost all of this energy has to go into
photon radiation and relativistic electron-positron pairs. This is
consistent with the neutrino heating processes. In absorption of $\nu_e$
and $\bar\nu_e$, essentially all the neutrino energy goes into the produced
electron or positron. Neutrino-antineutrino annihilation produces
electron-positron pairs, and neutrino-electron scattering also transfers
neutrino energy directly to electrons and positrons. Therefore, the ejecta
become radiation-dominated a short distance above the neutron star. 
The energy initially stored
in photon radiation and electron-positron pairs is converted into
the mechanical energy of the nucleons at much larger radii where
temperatures are low.

In terms of the local thermodynamic conditions, the dominance of 
radiation means that $T^3\gg\rho/m_N$ and $\eta\ll 1$. This is clear 
from equations (4)
to (6). Under these conditions, it is convenient to introduce a thermodynamic
quantity
\begin{equation}
S={11\pi^2\over 45}{T^3\over(\rho/m_N)}\approx 5.21{T_{\rm MeV}^3\over\rho_8},
\end{equation}
where $T_{\rm MeV}$ is the temperature in MeV, and $\rho_8$ is the density in
$10^8$ g cm$^{-3}$. It is easy to see from equations (4) to (6) that $S$ is the
entropy per baryon in relativistic particles for $\eta=0$. The ejecta 
become radiation-dominated when $S\gg1$.

\subsection{Input neutrino physics}

We now calculate the heating and cooling rates resulting from interactions
between the neutrino flux and material in the ejecta.  We assume that
neutrinos are emitted from a neutrinosphere with radius $R_\nu$. At radius
$r > R_\nu$, one only sees neutrinos within the
solid angle subtended by the neutrinosphere at this radius.  Because the
neutrino interaction cross sections have a power law dependence on neutrino
energy, the heating rates can be cast in terms of the neutrino luminosity and
various neutrino energy moments, without specifying a particular neutrino
energy distribution. Our approach here thus parallels pioneering analytic
calculations of the supernova mechanism by Bethe (e.g., 1993).

The most important heating and cooling processes are those given in equations
(8a) and (8b), i.e., neutrino absorption and electron capture on free nucleons.
The specific heating rate due to neutrino absorption is
\begin{eqnarray}
\dot q_{\nu N} &=&\dot q_{\nu_e n}+\dot q_{\bar\nu_e p}\nonumber\\
               &\approx& 9.65 N_A
\left[(1-Y_e)L_{\nu_e,51}\varepsilon^2_{\nu_e,\rm MeV}+
Y_e L_{\bar\nu_e,51}\varepsilon^2_{\bar\nu_e,\rm MeV}\right]
{1-x\over R_{\nu6}^2}
\ {\rm MeV}\ {\rm s}^{-1}\ {\rm g}^{-1},
\end{eqnarray}
where $x=\sqrt{1-R_\nu^2/r^2}$ is a function of radius, $N_A$ is the Avogadro's
number, $L_{\nu,51}$ is the individual neutrino luminosity in $10^{51}$ erg 
s$^{-1}$, $R_{\nu6}$ is the neutrinosphere radius in $10^6$ cm, and 
$\varepsilon_{\nu,\rm MeV}$ is an appropriate neutrino energy 
$\varepsilon_\nu$ in MeV, defined
through $\varepsilon^2_\nu\equiv\langle E_\nu^3\rangle/\langle E_\nu\rangle$,
where $\langle E_\nu^n\rangle$ denotes the $n$th neutrino energy moment of the
neutrino energy distribution. The specific cooling rate due to capture of
relativistic electrons and positrons is
\begin{equation}
\dot q_{eN}=\dot q_{e^-p}+\dot q_{e^+n}\approx 2.27 N_A T_{\rm MeV}^6\ 
{\rm MeV}\ {\rm s}^{-1}\ {\rm g}^{-1}.
\end{equation}
In deriving the above rates, we have neglected the neutron-proton mass
difference and the Pauli blocking effects for leptons in the final state.
We have also used the fact that there are almost equal numbers of 
relativistic electrons
and positrons in the radiation-dominated ejecta.

Neutrinos of all flavors can scatter on the electrons and positrons in the
ejecta. These scattering processes also contribute to the heating. The
corresponding specific heating rate is
\begin{eqnarray}
\dot q_{\nu e} &=& \dot q_{\nu_ee^-}+\dot q_{\nu_ee^+}+\dot q_{\bar\nu_ee^-}
+\dot q_{\bar\nu_ee^+}+2(\dot q_{\nu_\mu e^-}+\dot q_{\nu_\mu e^+}
+\dot q_{\bar\nu_\mu e^-}+\dot q_{\bar\nu_\mu e^+})\nonumber\\
&\approx& 2.17 N_A \ {\rm MeV}\ {\rm s}^{-1}\ {\rm g}^{-1}\nonumber\\
&\times& {T_{\rm MeV}^4\over\rho_8}\left(
L_{\nu_e,51}\epsilon_{\nu_e,\rm MeV}+
L_{\bar\nu_e,51}\epsilon_{\bar\nu_e,\rm MeV}+
{6\over 7}L_{\nu_\mu,51}\epsilon_{\nu_\mu,\rm MeV}\right)
{1-x\over R_{\nu6}^2},
\end{eqnarray}
where $\epsilon_{\nu,\rm MeV}$ is $\epsilon_\nu\equiv\langle E_\nu^2\rangle/
\langle E_\nu\rangle$ in MeV, and where we have assumed the same 
characteristics for the $\nu_\mu$, $\bar\nu_\mu$, $\nu_\tau$, and 
$\bar\nu_\tau$ fluxes. In deriving equation (12), we have also made the
approximation that the average energy transfer from neutrino to electron
is $E_\nu/2$ for each scattering, instead of the more accurate value
$(E_\nu-4T)/2$ found in Tubbs and Schramm (1975). This is a reasonable
approximation at $T\lesssim$ a few MeV where the neutrino-electron scattering
processes are expected to be most effective.

Cooling can also occur through annihilation of relativistic electron-positron
pairs into $\nu_e\bar\nu_e$, $\nu_\mu\bar\nu_\mu$, and $\nu_\tau\bar\nu_\tau$
pairs. The corresponding specific cooling rate is
\begin{equation}
\dot q_{e^-e^+}\approx 0.144 N_A {T^9_{\rm MeV}\over\rho_8}\ 
{\rm MeV}\ {\rm s}^{-1}\ {\rm g}^{-1}.
\end{equation}
Of course, there are other cooling processes operating through emission of
neutrino-antineutrino pairs. However, electron-positron pair annihilation
into neutrino-antineutrino pairs is the most important one for the
radiation-dominated conditions in the ejecta.

Finally, we give the specific heating rate for neutrino-antineutrino
annihilation into electron-positron pairs:
\begin{eqnarray}
\dot q_{\nu\bar\nu}&\approx&12.0 N_A\ {\rm MeV}\ {\rm s}^{-1}\ {\rm g}^{-1}
\nonumber\\
&\times&\left[L_{\nu_e,51}L_{\bar\nu_e,51}\left(\epsilon_{\nu_e,\rm MeV}
+\epsilon_{\bar\nu_e,\rm MeV}\right)+{6\over 7}L_{\nu_\mu,51}^2
\epsilon_{\nu_\mu,\rm MeV}\right]{\Phi(x)\over\rho_8R_{\nu6}^4},
\end{eqnarray}
where $\Phi(x)=(1-x)^4(x^2+4x+5)$. It is also useful to give the total
heating rate due to neutrino-antineutrino annihilation above the
neutrinosphere:
\begin{equation}
\dot Q_{\nu\bar\nu}\approx 4.85\times 10^{45}\left[L_{\nu_e,51}L_{\bar\nu_e,51}
\left(\epsilon_{\nu_e,\rm MeV}+\epsilon_{\bar\nu_e,\rm MeV}\right)+{6\over 7}
L_{\nu_\mu,51}^2\epsilon_{\nu_\mu,\rm MeV}\right]R_{\nu6}^{-1}\ {\rm erg}
\ {\rm s}^{-1}.
\end{equation}

The heating and cooling rates presented above correspond to the major heating
and cooling mechanisms for the ejecta. From these rates,
we can make the following
observations:

(1) The specific heating rate due to neutrino-antineutrino annihilation 
decreases rapidly away from the neutrinosphere, as is evident from the
expression for $\Phi(x)$. At $r\gg R_\nu$, $\Phi(x)\rightarrow 
(5/8)R_\nu^8/r^8$. Physically, this is because (i) the neutrino and 
antineutrino fluxes decrease
as $r^{-2}$; (ii) the chance for a neutrino
to collide with an antineutrino, especially the occurrence of almost head-on 
collisions
decreases quickly away from the neutrinosphere; and (iii) the center-of-mass
energy of the neutrino-antineutrino pair, and hence, the available phase space
for the produced electron-positron pair, also decrease very fast away from the 
neutrinosphere.
Unfortunately, our prescription for the neutrino flux does not work well
close to the neutrinosphere. Therefore, the rates given for this process
can only be taken as a rough estimate. Extensive studies of the heating rate
due to neutrino-antineutrino annihilation were carried out by Janka (1991a,b)
using Monte Carlo methods. In the subsequent (sub)sections, we will
discuss to what extent the simple approximations made in deriving equations (14)
and (15) will affect our understanding of the 
neutrino-driven wind.

(2) The relative importance of neutrino-scattering and neutrino absorption
can be seen from the ratio $\dot q_{\nu e}/\dot q_{\nu N}\sim (S/23)(T/T_\nu)$,
where $T_\nu\sim 5$ MeV is the temperature at the neutrinosphere. Therefore,
neutrino-electron scattering become important only at high entropies.

(3) Similarly, the relative importance of the cooling processes can be gauged
by the ratio $\dot q_{e^-e^+}/\dot q_{eN}\approx S/82$. However, both cooling
processes have sensitive temperature dependence, and the absolute cooling
rates decrease rapidly away from the neutron star surface.

\subsection{Initial conditions}
We refer to the conditions at radius $R$, where we start our analytic
treatment of the neutrino-driven wind, as the initial conditions
at the neutron star surface. These conditions are closely related to the
heating and cooling processes taking place in the region between the
neutrinosphere at radius $R_\nu$ and the neutron star surface at radius $R$.
In this
region, the heating and cooling processes between the intense neutrino flux
and material at high temperature and density proceed at very high rates.
This results in a kinetic equilibrium between the neutrino flux and the 
material (Burrows \& Mazurek 1982), and the temperature in this region 
almost stays constant.

The material in this region is also in close hydrostatic equilibrium, and
we have
\begin{equation}
-{1\over\rho}{dP\over dr}\approx{GM\over r^2}.
\end{equation}
For slowly varying temperature,
\begin{equation}
{dP\over dr}\approx{\rho T_\nu\over m_N}\left(Y_e{d\eta\over dr}+{1\over\rho}
{d\rho\over dr}\right).
\end{equation}
Because previous deleptonization has led to $Y_e\ll 0.5$ at the 
neutrinosphere, we can neglect the first term in the above equation, and obtain
\begin{equation}
\rho(r)\approx\rho(R_\nu)\exp\left(-{GMm_N\over R_\nu T_\nu}{r-R_\nu\over 
R_\nu}\right)     
\end{equation}
for $r-R_\nu\ll R_\nu$. Taking $M\sim 1.4 M_{\sun}$, $R_\nu\sim 10$ km, and
$T_\nu\sim 5$ MeV, we see that the density decreases by one $e$-fold over a
distance scale $\sim R_\nu^2T_\nu/GMm_N\sim 0.25$ km. As a result, the entropy
in relativistic particles $S\propto T^3/\rho$ quickly rises.

The pressure and specific internal energy of the material become dominated by 
contributions from the relativistic particles when $S\gtrsim 4$. At this point,
the temperature has to decrease sufficiently fast to break the kinetic 
equilibrium in order to
maintain the approximate hydrostatic equilibrium. The cooling rates decrease
with the falling temperature, and net heating takes place to drive a mass
outflow, and to further increase the entropy. This picture is then consistent
with our expectation of the general radiation-dominated condition in the
ejecta.

Our initial conditions then correspond to the last radius where kinetic 
equilibrium 
can still be maintained. At this radius $R$, $S\gtrsim 4$, so the dominant
cooling process is electron capture on free nucleons. Since the rate for
this process has a sharp temperature dependence, the temperature $T_i$ at
radius $R$ is relatively insensitive to the uncertainties in the rates
for the counteracting heating processes. We can then take neutrino absorption
on free nucleons as the dominant heating process, and further approximate
the corresponding rate in equation (10) as
\begin{equation}
\dot q_{\nu N}\approx 2.75N_A (L_{\nu_e,51}\epsilon_{\nu_e,\rm MeV}^2+
L_{\bar\nu_e,51}\epsilon_{\bar\nu_e,\rm MeV}^2)r^{-2}_6
\ {\rm MeV}\ {\rm s}^{-1}\ {\rm g}^{-1}.
\end{equation}
In deriving the above rate, we have made the following approximations:
(i) $Y_e\approx 0.5$; (ii) $1-x\approx (1/2)R_\nu^2/r^2$; and (iii)
$\varepsilon_\nu^2\approx 1.14\epsilon_\nu^2$. The last approximation
follows from the observation that $\langle E_\nu^3\rangle\langle E_\nu\rangle/
\langle E_\nu^2\rangle^2$ ranges from 25/24 to 5/4 for neutrino energy
distributions of the form $f(E_\nu)\propto E_\nu^2/
[\exp(a E_\nu+b)+1]$ with $a>0$ and $-\infty<b<\infty$. The
temperature $T_i$ at radius $R$ is then
\begin{equation}
T_i\approx 1.03\left[1+{L_{\nu_e}\over L_{\bar\nu_e}}\left({\epsilon_{\nu_e}
\over \epsilon_{\bar\nu_e}}\right)^2\right]^{1/6}{L_{\bar\nu_e,51}^{1/6}
\over R_6^{1/3}}\epsilon_{\bar\nu_e,\rm MeV}^{1/3} \ {\rm MeV}.
\end{equation}
For $L_{\bar\nu_e}\approx 10^{51}$ erg s$^{-1}$, $\epsilon_{\bar\nu_e}\approx
20$ MeV, $L_{\nu_e}/L_{\bar\nu_e}\approx 1$, $\epsilon_{\nu_e}/
\epsilon_{\bar\nu_e}\approx$ 0.5--1, and $R\approx 10^6$ cm, we have $T_i
\approx 3$ MeV.

We can now compare the rates $\dot q_{\nu e}$ and $\dot q_{\nu\bar\nu}$ with
$\dot q_{\nu N}$. As we have discussed in $\S3.1$, $\dot q_{\nu e}/
\dot q_{\nu N}\approx(S/23)(T_i/T_\nu)$. For $S\gtrsim 4$ and $T_i\lesssim 
T_\nu$, $\dot q_{\nu e}/\dot q_{\nu N}\approx 1/6$. Taking $L_{\nu_e}\approx
L_{\bar\nu_e}\approx L_{\nu_\mu}\approx 10^{51}$ erg s$^{-1}$ and $R\approx
R_\nu\approx 10^6$ cm, we obtain $\dot q_{\nu\bar\nu}/\dot q_{\nu e}\approx
27.6/T_{\rm MeV}^4\approx 1/3$. Clearly, taking into account the contributions
from neutrino-electron scattering and neutrino-antineutrino annihilation only
increases $T_i$ by a small amount. However, to guard against miscalculation
of the heating rates, especially $\dot q_{\nu\bar\nu}$, we can multiply
$\dot q_{\nu N}$ by a correction factor $C$. This then gives the temperature
$T_i$ at radius $R$ as
\begin{equation}
T_i\approx 1.11C^{1/6}L_{\bar\nu_e,51}^{1/6}
\epsilon_{\bar\nu_e,\rm MeV}^{1/3}R_6^{-{1/3}}\ {\rm MeV},
\end {equation}  
where we have assumed an average value for $[1+(L_{\nu_e}/L_{\bar\nu_e})
(\epsilon_{\nu_e}/\epsilon_{\bar\nu_e})^2]^{1/6}$. In the unlikely case where
we have underestimated $\dot q_{\nu\bar\nu}$ by more than one order of 
magnitude, we note that a correction factor $C=5$ only increases $T_i$ by
$\sim 30$\%. Equation (21) then constitutes our basic initial condition at
the neutron star surface.

\subsection{Analytic treatment}
Before we start our analytic treatment of the neutrino-driven wind,
we note that equations (1) to (3) describe the hydrodynamic and
thermodynamic evolution of the ejecta, while equation (7) determines the
chemical evolution of the ejecta. The former equations are coupled with
the latter only through the dependence on $Y_e$ in the equations of state
(4) to (6) and the expression of $\dot q$. From the previous discussions
in this section, we know that the ejecta become radiation-dominated
immediately above the neutron star surface. Equations (4) and (5) are
approximately reduced to
\begin{eqnarray}
P &\approx& {11\pi^2\over 180}T^4,\\
\epsilon &\approx& {11\pi^2\over 60}{T^4\over\rho},
\end{eqnarray}
which are independent of $Y_e$. The expression of $\dot q$ has a weak
dependence on $Y_e$, mainly through equation (10). We have eliminated this
dependence by taking $Y_e\approx 0.5$ in $\S3.2$. This is justified because
$(1-Y_e)L_{\nu_e}\varepsilon_{\nu_e}^2+Y_eL_{\bar\nu_e}
\varepsilon_{\bar\nu_e}^2\approx (1/2)(L_{\nu_e}\varepsilon_{\nu_e}^2+
L_{\bar\nu_e}\varepsilon_{\bar\nu_e}^2)$ for any $Y_e$ as long as $L_{\nu_e}
\approx L_{\bar\nu_e}$ and $\varepsilon_{\nu_e}\sim\varepsilon_{\bar\nu_e}$.
At any rate, we have introduced a correction factor $C$ in $\S3.2$ to account
for the uncertainties in deriving the expression of $\dot q$. Therefore, we
can separate the discussion of the ejecta into two almost independent parts.
In this subsection, we derive the analytic expressions for the entropy, the
dynamic time scale, and the mass outflow rate in the ejecta. These physical
parameters are determined by equations (1) to (3). We devote $\S5$ to a
detailed analysis of the electron fraction in the ejecta.

Taking advantage of equations (22) and (23), we can rearrange equations (1) to
(3) into
\begin{eqnarray}
\left(v-{v_s^2\over v}\right){dv\over dr}&=&{1\over r}\left({2\over 3}{TS\over 
m_N}-
{GM\over r}\right)-{\dot q\over 3v},\\
\dot q &=& v{d\over dr}\left({v^2\over 2}+{TS\over m_N}-{GM\over r}\right),\\
v{dS\over dr}&=&{\dot q m_N\over T},
\end{eqnarray}
where $v_s$ is the adiabatic sound speed in the ejecta, given by
\begin{equation}
v_s=\left({4\over 3}{P\over\rho}\right)^{1/2}=\left({TS\over 3m_N}\right)^{1/2}
\approx 5.67\times 
10^8T_{\rm MeV}^{1/2}S^{1/2}\ {\rm cm}\ {\rm s}^{-1}.
\end{equation}
For a particular Lagrangian mass element, the above equations describe the
evolution of its velocity, total ``flow'' energy, and entropy. We note that
the total flow energy is given by the sum of enthalpy and mechanical energy,
i.e., $\epsilon_{\rm flow}=v^2/2+TS/m_N-GM/r$.

Given the initial conditions at the neutron star surface, i.e., $T$ and $S$
at radius $R$, we can choose any three independent equations from equations
(1) to (3) and (24) to (26), and integrate these equations for various values
of the mass outflow rate $\dot M$. The eigenvalue of $\dot M$ is then selected
to meet the boundary conditions at the shock wave. However, the qualitative
features of the solutions to these equations can be readily obtained from
equations (24) to (26).

From equations (25) and (26), we see that $\epsilon_{\rm flow}$ and $S$
always increase as long as $\dot q>0$. Neutrino heating becomes negligible
for $T\lesssim0.5$ MeV because in the region at these temperatures, (i)
free nucleons are bound into $\alpha$-particles and heavier nuclei; 
(ii) electron-positron pairs
annihilate into photons; and (iii) the neutrino flux has decreased by a
factor $(R_\nu/r)^2$. Therefore, $\epsilon_{\rm flow}$ and $S$ reach their
final values at $T\approx 0.5$ MeV. We note that equations (22) and (23) are
invalid for $T\lesssim 0.5$ MeV due to the disappearance of electron-positron
pairs. However, because entropy is conserved during the annihilation of 
electron-positron pairs into photons, we can show that after all the
electron-positron pairs are gone, the correct equations to describe the ejecta
are
\begin{eqnarray}
\left(v-{v_s^2\over v}\right){dv\over dr} &=& {1\over r}\left({2\over 3}
{TS_f\over m_N}-{GM\over r}\right),\\
\epsilon_{{\rm flow},f} &=& \left({v^2\over 2}+{TS_f\over m_N}-{GM\over r}
\right),\\
S_f &=& {4\pi^2\over 45}{T^3\over \rho},
\end{eqnarray}
where $\epsilon_{{\rm flow},f}$ and $S_f$ are the final values of 
$\epsilon_{\rm flow}$
and $S$ at $T\approx 0.5$ MeV, respectively, and $v_s=(TS_f/3m_N)^{1/2}$.

The behavior of $v$ is more complicated. At the neutron star surface, we have
$v\ll v_s$ and $\epsilon_{\rm flow}\approx -GM/R$. According to equation (24),
$v$ always increases initially. The subsequent evolution of $v$ depends on the
mass outflow rate $\dot M$. Given the initial conditions at the neutron star
surface, the initial velocity is proportional to $\dot M$. For small values
of $\dot M$, the ejecta move slowly and there is more time for the ejecta to
gain energy from the neutrino flux. The right-hand side of equation (24) can
become zero due to the increase in $\epsilon_{\rm flow}$ before $v$ reaches
$v_s$. When this occurs, $dv/dr=0$, and $v$ reaches its maximum value. The
velocity decreases afterwards as the right-hand side of equation (24) becomes
positive. In this case, the ejecta are always subsonic. When $\dot M$ 
increases, it can happen that the right-hand side of equation (24) becomes
zero at the same time as $v$ reaches $v_s$. This case corresponds to a
critical value $\dot M_{\rm crit}$ of $\dot M$. In this critical case, the
velocity increases through the sound speed to supersonic values, and eventually
becomes a constant when all the flow energy is converted into the mechanical
kinetic energy. Mass outflow rates larger than $\dot M_{\rm crit}$ are 
unphysical,
because for these values of $\dot M$, $v$ will reach $v_s$ when the right-hand 
side of equation (24) is
still negative, resulting in an unphysical infinite acceleration. Therefore,
we only need to focus our attention on cases of $\dot M\leq \dot M_{\rm crit}$.

From equations (1) and (30), we see that $\rho\propto r^{-2}$ and $T\propto
r^{-2/3}$ as $v$ approaches its final value in the critical case. Therefore,
$\dot M_{\rm crit}$ can only correspond to vanishing boundary conditions at
large radii. On the other hand, in the cases of $\dot M<\dot M_{\rm crit}$, one
can impose a boundary pressure, or equivalently a boundary temperature, in
the radiation-dominated ejecta. The supernova shock wave at large radii can
be regarded as such a boundary condition. As $T$ in the ejecta approaches
the boundary temperature $T_b$, $\rho$ also approaches a constant value,
and $v$ decreases as $r^{-2}$, according to equations (1) and (30). However,
for $T_b\ll 0.5$ MeV, there is little difference between the subsonic
structures of the ejecta for the critical and subcritical mass outflow rates.

In the subcritical case, $v<v_s$ everywhere, and we can approximate equation
(29) as
\begin{equation}
{TS_f\over m_N}-{GM\over r}\approx \epsilon_{{\rm flow},f}\approx 
{T_bS_f\over m_N}.
\end{equation}
In particular, this equation is also satisfied at $T\approx 0.5$ MeV because
$\epsilon_{{\rm flow},f}$ corresponds to the flow energy at $T\approx 0.5$ MeV.
If $T_b\ll 0.5$ MeV, we can take
\begin{equation}
{TS_f\over m_N}\approx {GM\over r}\ {\rm at}\ T\approx0.5\ {\rm MeV}
\end{equation}
as an effective boundary condition. In the critical case, we have
$(2/3)TS_f/m_N=GM/r$ when $v=v_s$. Equation (29) then gives
\begin{equation} 
{v^2\over 2}+{TS_f\over m_N}-{GM\over r}=\epsilon_{{\rm flow},f}\approx {T_sS_f
\over 2m_N},
\end{equation}
where $T_s$ is the temperature when $v$ reaches $v_s$. If $T_s\ll 0.5$ MeV,
i.e., $T\approx 0.5$ MeV belongs to the subsonic region, we obtain the same 
effective boundary condition as in equation (32). Therefore, the mass outflow 
rates are close to $\dot M_{\rm crit}$ for small boundary temperatures,
resulting in similar subsonic structures to that in the critical case.
Typical boundary temperatures
for the supernova shock wave are $T_b\sim 0.1 $ MeV. Therefore, the mass
outflow rate in the neutrino-driven wind corresponds to the
effective boundary condition in equation (32). As we have stated before, our
discussion of the physical conditions in the ejecta can be limited to the 
region at $T\gtrsim 0.5$ MeV.

The above discussion of the neutrino-driven wind is mainly based
on equations (24) to (26). Strictly speaking, these equations are valid only
when $S\gg S_N$, where
\begin{equation}
S_N\approx 11+\ln\left({T_{\rm MeV}^{3/2}\over \rho_8}\right)
\end{equation}
is the entropy per baryon in non-relativistic nucleons. The number 11 in the
above expression for $S_N$ could be 12 if one distinguishes between neutrons
and protons, and takes account of their spin states. However, the qualitative
features given in the above discussion apply to the neutrino-driven wind
as long as the pressure $P$ and the specific energy $\epsilon$ are
dominated by the contributions from electron-positron pairs and photon
radiation, i.e., $S>4$. In this case, the analytic treatments of the
neutrino-driven wind are very similar for both cases of
$S\gg S_N$ and $S\lesssim S_N$, if we use the general forms of equations
(25) and (26):
\begin{eqnarray}
\dot q &=& v{d\over dr}\left({v^2\over 2}+{TS\over m_N}+{5\over 2}{T\over m_N}
-{GM\over r}\right),\\
{\dot qm_N\over T} &=& v{d\over dr}\left(S+S_N\right).
\end{eqnarray} 
Equations (35) and (36) can be easily derived from equations (2) to (5) with
the approximation $\eta\ll 1$.

\subsubsection{Entropy per baryon}

We first derive the final entropy $S_f$ in the ejecta. From equation (35),
we have
\begin{equation} 
{GM\over R}\approx \epsilon_{{\rm flow},f}-\epsilon_{{\rm flow},i}
=\int_R\dot q{dr\over v},
\end{equation}
where we have used the effective boundary condition in equation (32) and have
taken the initial flow energy to be $\epsilon_{{\rm flow},i}\approx -GM/R$.
From equation (36), we have
\begin{mathletters}
\begin{eqnarray}
S_f &\approx& \int_R{\dot qm_N\over T}{dr\over v}
\ {\rm for}\ S_f\gg S_N,\\
S_f &\approx& {1\over 2}\int_R{\dot qm_N\over T}{dr\over v}
\ {\rm for}\ 
S_f\lesssim S_N.
\end{eqnarray}
\end{mathletters}
We can define an effective temperature $T_{\rm eff}$ through
\begin{equation}
T_{\rm eff}^{-1}={\int_R(\dot q/T)(dr/v)\over\int_R\dot q(dr/v)}
={\int_R(\dot q/T)dt\over
\int_R\dot qdt},
\end{equation}
or equivalently,
\begin{mathletters}
\begin{eqnarray}
S_f &\approx& {GMm_N\over RT_{\rm eff}}\ {\rm for}\ S_f\gg S_N,\\
S_f &\approx& {GMm_N\over 2RT_{\rm eff}}\ {\rm for}\ S_f\lesssim S_N.
\end{eqnarray}
\end{mathletters}
We can regard equation (39) as defining $T_{\rm eff}^{-1}$ by weighing 
$T^{-1}$ over $\dot q$, and expect $T_{\rm eff}$ approximately corresponds
to the temperature at which $\dot q$ reaches the maximum value. 

Taking
$\dot q\approx C\dot q_{\nu N}-\dot q_{eN}\propto T_i^6(R/r)^2-T^6$, we find
\begin{equation}
T^5{dT\over dr}=-{1\over 3}T_i^6{R^2\over r^3}
\end{equation}
at $T_{\rm eff}$, where we have used the same correction factor introduced in
equation (21) to crudely account for the heating processes other than neutrino
absorption on free nucleons. We note that in general this correction factor
$C$ should vary with radius, for example, due to the increasing importance
of $\dot q_{\nu e}$ relative to $\dot q_{\nu N}$ at higher entropies. However,
for the purpose of simple analytic treatments, we will take $C$ to be constant
and subsequently derive its suitable value to be used in the analytic estimates
for the entropy per baryon, the dynamic time scale, and the mass outflow rate
in the wind.

Because hydrostatic equilibrium approximately holds in
the subsonic region, we have
\begin{equation}
{1\over \rho}{dP\over dr}=\beta{S\over m_N}{dT\over dr}\approx -{GM\over r^2},
\end{equation}
where we have introduced a numerical factor $\beta$ to account for the
contribution to $P$ from non-relativistic free nucleons. For $S_f\gg S_N$,
$\beta\approx 1$, and for $S_f\lesssim S_N$, $\beta\approx 2$.
Equations (41) and (42) then give
\begin{equation}
T_{\rm eff}=T_i\left({\beta\over 3}{T_{\rm eff}S_{\rm eff}r_{\rm eff}\over 
GMm_N}
\right)^{1/6}
\left({R\over r_{\rm eff}}\right)^{1/3},
\end{equation}
where $r_{\rm eff}$ and $S_{\rm eff}$ are the radius and entropy, respectively,
corresponding to $T_{\rm eff}$. Hereafter,
we use the subscript ``eff'' to denote parameters at the radius where $\dot q$
reaches the maximum value. To proceed further, we make the following
approximation:
\begin{equation}
S_{\rm eff}\approx {S_f\over 2},
\end{equation}
and
\begin{equation}
T_{\rm eff}S_{\rm eff}\approx {GMm_N\over 2r_{\rm eff}}.
\end{equation}
The numerical factors in equations (44) and (45) are conveniently chosen
to reflect that the concerned quantities at $r_{\rm eff}$ are in the
``middle'' of some characteristic evolution.
Using equations (40a), (40b), (44), and (45), we find
\begin{mathletters}
\begin{eqnarray}
r_{\rm eff} &\approx& R\ {\rm for}\ S_f\gg S_N,\\
r_{\rm eff} &\approx& 2R\ {\rm for}\ S_f\lesssim S_N.
\end{eqnarray}
\end{mathletters}
Substituting equations (45), (46a), and (46b) into equation (43), we have
\begin{mathletters}
\begin{eqnarray}
T_{\rm eff} &\approx& 6^{-1/6}T_i\ {\rm for}\ S_f\gg S_N,\\
T_{\rm eff} &\approx& 12^{-1/6}T_i\ {\rm for}\ S_f\lesssim S_N.
\end{eqnarray}
\end{mathletters}

Combining equations (21), (40a), (40b), (47a), and (47b), we obtain
\begin{mathletters}
\begin{eqnarray}
S_f &\approx& 235C^{-1/6}L_{\bar\nu_e,51}^{-1/6}
\epsilon_{\bar\nu_e,\rm MeV}^{-1/3}R_6^{-2/3}\left({M\over 1.4 M_{\sun}}\right)
\ {\rm for}\ S_f\gg S_N,\\
S_f &\approx& 132C^{-1/6}L_{\bar\nu_e,51}^{-1/6}
\epsilon_{\bar\nu_e,\rm MeV}^{-1/3}R_6^{-2/3}\left({M\over 1.4 M_{\sun}}\right)
\ {\rm for}\ S_f\lesssim S_N.
\end{eqnarray}
\end{mathletters}
The total entropy per baryon $S_{\rm tot}$ in the ejecta is
\begin{equation}
S_{\rm tot}\approx S_f+S_N\approx S_f+\ln S_f+10,
\end{equation}
where we have evaluated $S_N$ at $T\approx 0.5$ MeV using $S_f$ and taking
advantage of the logarithmic dependence of $S_N$ on $T$.

We can also derive an approximate value for the correction factor $C$. Assuming
that the main heating process other than neutrino absorption on free nucleons
is neutrino-electron scattering, we can take the correction factor to be
\begin{equation}
C\approx 1+{\dot q_{\nu e,\rm eff}\over\dot q_{\nu N,\rm eff}}.
\end{equation}
From equations (12) and (19), we obtain
\begin{mathletters}
\begin{eqnarray}
{\dot q_{\nu e,\rm eff}\over\dot q_{\nu N,\rm eff}} &\approx&
7.33R_6^{-1}\left({M\over 1.4 M_{\sun}}\right){\epsilon_{\nu_e,\rm MeV}+
\epsilon_{\bar\nu_e,\rm MeV}+(6/7)\epsilon_{\nu_\mu,\rm MeV}\over
\epsilon_{\nu_e,\rm MeV}^2+\epsilon_{\bar\nu_e,\rm MeV}^2}\ {\rm for}\ 
S_f\gg S_N,\\
{\dot q_{\nu e,\rm eff}\over\dot q_{\nu N,\rm eff}} &\approx&
3.66R_6^{-1}\left({M\over 1.4 M_{\sun}}\right){\epsilon_{\nu_e,\rm MeV}+
\epsilon_{\bar\nu_e,\rm MeV}+(6/7)\epsilon_{\nu_\mu,\rm MeV}\over
\epsilon_{\nu_e,\rm MeV}^2+\epsilon_{\bar\nu_e,\rm MeV}^2}\ {\rm for}\ 
S_f\lesssim S_N,
\end{eqnarray}
\end{mathletters}
where we have used equations (45), (46a), and (46b) to evaluate $T^4/\rho$ in
equation (12), and where we have assumed $L_{\nu_e}\approx L_{\bar\nu_e}
\approx L_{\nu_\mu}$. Typically, we have $\epsilon_{\nu_e}+\epsilon_{\bar\nu_e}
\approx \epsilon_{\nu_\mu}$, $\epsilon_{\nu_e}\approx$ 
(0.5--1)$\epsilon_{\bar\nu_e}$, and $\epsilon_{\bar\nu_e}\approx 20$ MeV. So
numerically, $[\epsilon_{\nu_e,\rm MeV}+\epsilon_{\bar\nu_e,\rm MeV}+
(6/7)\epsilon_{\nu_\mu,\rm MeV}]/
(\epsilon_{\nu_e,\rm MeV}^2+\epsilon_{\bar\nu_e,\rm MeV}^2)\approx 0.1.$

\subsubsection{Mass outflow rate}

Next, we derive the mass outflow rate $\dot M$ in the ejecta.
The total net neutrino heating rate above $r_{\rm eff}$ is given by
\begin{equation}
\int_{r_{\rm eff}}\dot q{dr\over v}\approx C\int_{r_{\rm eff}}\dot q_{\nu N,i}
{R^2\over r^2}{dr\over v}\approx C\dot q_{\nu N,i}{4\pi R^2\over \dot M}
\int_{r_{\rm eff}}\rho dr,
\end{equation}
where
\begin{equation}
C\dot q_{\nu N,i}\approx 2.27N_AT_{i,\rm MeV}^6\ {\rm MeV}\ {\rm s}^{-1}\ 
{\rm g}^{-1}\approx 2.19\times 10^{18}T_{i,\rm MeV}^6\ {\rm erg}\ 
{\rm s}^{-1}\ {\rm g}^{-1}.
\end{equation}
To estimate the integral in equation (52), we need the density scale height
$h_{\rho,\rm eff}$ at $r_{\rm eff}$. We can relate $h_{\rho,\rm eff}$ to
the corresponding temperature scale height $h_{T,\rm eff}$ by
\begin{equation}
h_{\rho,\rm eff}\approx  h_{T,\rm eff}/3,
\end{equation}
for the radiation-dominated ejecta ($S\gtrsim 4$). In turn, 
$h_{T,\rm eff}$ can be 
estimated from equation (42) as
\begin{equation}
h_{T,\rm eff}=\left|{d\ln T\over dr}\right|^{-1}_{\rm eff}\approx
\beta{r_{\rm eff}^2T_{\rm eff}S_{\rm eff}\over GMm_N}.
\end{equation}
Now the integral in equation (52) can be approximated as
\begin{equation}
\int_{r_{\rm eff}}\rho dr\approx \rho_{\rm eff}h_{\rho,\rm eff}\approx
{11\pi^2\over 135}\beta{T_{\rm eff}^4r_{\rm eff}^2\over GM},
\end{equation}
where we have used equation (9) to express $\rho_{\rm eff}$ in terms of
$T_{\rm eff}$ and $S_{\rm eff}$.

From equation (35), we have
\begin{equation}
\epsilon_{{\rm flow},f}-\epsilon_{\rm flow,eff}\approx {1\over 2}{GM\over
r_{\rm eff}}\approx\int_{r_{\rm eff}}\dot q{dr\over v},
\end{equation}
where we have used equation (45) to evaluate $\epsilon_{\rm flow,eff}$.
Combining equations (52), (53), (56), and (57), we obtain
\begin{mathletters}
\begin{eqnarray}
\dot M &\approx& 1.14\times 10^{-10}C^{5/3}L_{\bar\nu_e,51}^{5/3}
\epsilon_{\bar\nu_e,\rm MeV}^{10/3}R_6^{5/3}\left({1.4M_{\sun}\over M}\right)^2
\ M_{\sun}\ {\rm s}^{-1},\ {\rm for}\ S_f\gg S_N,\\
\dot M &\approx& 1.15\times 10^{-9}C^{5/3}L_{\bar\nu_e,51}^{5/3}
\epsilon_{\bar\nu_e,\rm MeV}^{10/3}R_6^{5/3}\left({1.4M_{\sun}\over
M}\right)^2 \ M_{\sun}\ {\rm s}^{-1},\ {\rm for}\ S_f\lesssim S_N,
\end{eqnarray}
\end{mathletters}
where we have also used equations (21), (46a), (46b), (47a), and (47b). The
total power of net neutrino heating $\dot Q\approx GM\dot M/R$ is approximately
given by
\begin{mathletters}
\begin{eqnarray}
\dot Q &\approx& 4.25\times 10^{43}C^{5/3}L_{\bar\nu_e,51}^{5/3}
\epsilon_{\bar\nu_e,\rm MeV}^{10/3}R_6^{2/3}\left({1.4M_{\sun}\over
M}\right) \ {\rm erg}\ {\rm s}^{-1}\ {\rm for}\ S_f\gg S_N,\\
\dot Q &\approx& 4.28\times 10^{44}C^{5/3}L_{\bar\nu_e,51}^{5/3}
\epsilon_{\bar\nu_e,\rm MeV}^{10/3}R_6^{2/3}\left({1.4 M_{\sun}\over
M}\right) \ {\rm erg}\ {\rm s}^{-1}\ {\rm for}\ S_f\lesssim S_N.
\end{eqnarray}
\end{mathletters}

\subsubsection{Dynamic time scale}
Finally, we derive the dynamic time scale in the ejecta. We define this time
scale to be 
\begin{equation}
\tau_{\rm dyn}\equiv\left.{r\over v}\right|_{T\approx0.5\ {\rm MeV}}.
\end{equation}
From equations (1), (9), (32), (48a), (48b), (58a), and (58b), we find
\begin{equation}
\tau_{\rm dyn}\approx 68.4C^{-1}L_{\bar\nu_e,51}^{-1}
\epsilon_{\bar\nu_e,\rm MeV}^{-2}
R_6\left({M\over 1.4M_{\sun}}\right)\ {\rm s}
\end{equation}
for both cases of $S_f\gg S_N$ and $S_f\lesssim S_N$. The time scale derived
in equation (61) will prove important for heavy element nucleosynthesis in
the neutrino-driven wind. We illustrate its usefulness as follows.

We expect $GMm_N/r\approx TS_f$, or equivalently $T\propto r^{-1}$, to 
approximately hold between $T\approx 0.5$ MeV and $T\gtrsim T_b\sim 0.1$ MeV.
If we ignore the effect of annihilation of electron-positron pairs into
photons, we roughly have $S_f\propto T^3/\rho$ and $\rho\propto r^{-3}$,
which leads to $v\propto r$ for a constant mass outflow rate $\dot M=4\pi
r^2\rho v$. Therefore, the time for the temperature to decrease from $T\approx
0.5$ MeV to, say 0.2 MeV, is $\sim\tau_{\rm dyn}\ln(0.5/0.2)\sim\tau_{\rm dyn}$.

Between the radius where the temperature reaches $T\approx T_b\sim 0.1$ MeV
for the first time and the position of the shock wave, both the temperature
and density of the ejecta remain approximately constant. According to
$\dot M=4\pi r^2\rho v$, we have $v\propto r^{-2}$, from which the time
dependence of the radius for a particular Lagrangian mass element can be
obtained with $r/v\sim\tau_{\rm dyn}$ at the initial radius for $T\approx T_b$.

To summarize, we have derived
the total entropy per baryon, the mass outflow rate, and the relevant
dynamic time scale for heavy element nucleosynthesis in the neutrino-driven
wind.
In particular, we have shown that these physical conditions
for heavy element nucleosynthesis only depend on the neutrino luminosity and
energy distributions, and the mass and radius of the neutron star.

\section{Numerical runs using KEPLER}

In order to test some of the approximations derived in the previous section,
a series of calculations were carried out using the one-dimensional implicit
hydrodynamic code, KEPLER (Weaver, Zimmerman, \& Woosley 1978). This
Lagrangian-coordinate-based code has an appropriate equation of state for the
ions, radiation, electrons, and pairs under all conditions of interest here.
For example, this equation of state adequately treats electrons
in any arbitrarily relativistic and degenerate regime.
It is most important that the code is implicitly differenced and thus capable
of accurately following the evolution in a situation where, as here, tight
hydrostatic equilibrium exists. Though the code has provision for radiation
transport, energy transport other than by neutrinos is negligible here. Only
the outer layers of the neutron star were carried, approximately 0.01
$M_{\sun}$. The remainder of the neutron star was represented by a boundary
condition of given radius, gravitational potential, and neutrino luminosity.
Typical densities and temperatures at the basis of the problem were 10$^{13}$
g cm$^{-3}$ and 8--10 MeV. The wind of course originated at much lower
densities and temperatures in a small amount of mass.

Also important was the code's ability to smoothly and repeatedly rezone the
material in the wind as its density decreased. Typical zoning in the wind
was about $10^{-8} M_{\sun}$. The velocity structure and, in the asymptotic
region, the mass outflow rate, $\dot M=4 \pi r^2 \rho v$, were, at all times
following a start-up transient, well behaved and smooth.

On the other hand, the neutrino physics employed was primitive by the
standard of modern supernova codes. The inner boundary acted as a ``light
bulb" and the overlying layers were assumed to subtract nothing from the
flux, i.e., the optically thin limit was assumed. The neutrino luminosity was
shared equally among the six varieties and each neutrino species was 
characterized by
its mean energy and flux. For a neutrinosphere radius
of $R_\nu=10$ km (also the approximate neutron star radius, $R$), the 
assumed
mean energies were $\epsilon_\nu\equiv
\langle E_\nu^2\rangle/\langle E_\nu\rangle$ =  
12, 22, 34 MeV 
for $\nu_e$, 
$\bar \nu_e$, and
$\nu_{\mu}$, respectively. The corresponding values for $R_\nu=30$ km were 16, 
20, 32
MeV. These neutrinos could interact with nucleons and pairs (and with one
another) in the neutron star crust and wind according to equations
(10), (11), (12), and (14). Neutrino losses due to pair annihilation, plus
plasma and photoneutrino processes were calculated separately assuming $Y_e
\approx 0.50$ and the fitting functions of  Munakata, Koyama, \& Itoh (1985).
This is more accurate than equation (13).

Nuclear physics, other than the neutrino interactions, was very simple.
Since in the region of interest, nuclear statistical equilibrium (NSE) 
exists and
the composition is solely neutrons, protons, and $\alpha$-particles, only these
three constituents were carried. Since the neutrino interactions with the two
nucleon species are similar only one species was carried. The transition from
nucleons to $\alpha$-particles, which effectively shuts off the 
neutrino-nucleon
interaction (neutrino-electron scattering and neutrino-antineutrino 
annihilation still contribute) was
approximated using the following prescription for the free nucleon mass
fraction: 
\begin{equation}
X_N \approx 828 {T_{\rm
MeV}^{9/8}\over\rho_8^{3/4}} \exp (-7.074/T_{\rm MeV})
\end{equation} 
or unity, whichever was the smaller (Woosley \& Baron 1992).
The free nucleon mass fraction $X_N$ given by this prescription was used to 
multiply the rates in 
equations (10) and (11)
to give the capture rates of neutrinos and pairs on nucleons.

\subsection {Newtonian calculations}

Nine models were then calculated. In each model the neutron star was represented
by an inner boundary condition of given interior mass, radius, and neutrino
luminosity. The parameters employed as well as sample results are given in
Table 1 and Figs. 1 and 2.

Each model was constructed by allowing a coarsely-zoned version of the neutron
star crust and atmosphere to come into approximate hydrostatic equilibrium
with the inner boundary conditions. This typically took a few milliseconds. 
Then the
zoner was turned on. Both adzoning and dezoning were employed. This acted to
give smooth temperature and density profiles. Typically in the steady-state
solution there were about 200 zones and about a dozen zones per decade of
density in the region of interest. Typical zone masses were $10^{-8} M_{\sun}$
except near the inner boundary where the zoning smoothly increased on
a logarithmic scale to $\sim 10^{-3} M_{\sun}$. Zones were removed from the
surface of the problem, typically when the density had declined to about 1 g
cm$^{-3}$ so as to inhibit a runaway accumulation of zones.

Each problem was run long enough that a steady state had clearly been
attained. For total neutrino luminosities of 10$^{52}$ erg s$^{-1}$ this took
about one second. For higher luminosities, the corresponding time was about
0.2 s. Here we point out that our numerical approach to model the 
neutrino-driven wind differs from that of Duncan et al. (1986). Their numerical
model of the neutrino-driven wind was based on the wind equations (1) to (3),
and mathematically was an eigenvalue problem looking for the mass outflow
rate $\dot M$. In our approach, we started from some initial configuration
of the material above the neutron star surface, and evolved the mass outflow
into a steady-state neutrino-driven wind. Therefore, not only did we obtain
the mass outflow rate $\dot M$ and other physical parameters in the steady
state, but also showed that a steady state could actually be reached over
some relaxation time scale of the problem. The relaxation time scale could
then be compared with the evolution time scales of the inner boundary 
(especially the neutrino luminosity) to establish the quasi-steady-state
nature of the neutrino-driven wind. For typical decay time of $\sim$ several
seconds for the neutrino luminosity, the steady-state approximation in our
analytic treatments of the neutrino-driven wind is valid.

Because the code did not include general relativistic corrections to
gravity, two different baryon masses, 1.4 and 2.0 $M_{\sun}$, were employed
for the neutron star. The smaller value is more typical of the gravitational
mass of neutron stars and the baryon mass of many presupernova model
calculations (e.g., Timmes, Woosley, \& Weaver 1996), but the latter gives
a more realistic gravitational potential, at least near the neutron star
surface, when the neutron star has approximately reached its final radius
of about 10 km.

Table 1 compares numerical and analytic results for the total power of
net neutrino heating, the mass outflow rate, the final total entropy per
baryon, and the dynamic time scale in the neutrino-driven wind. As we can
see, the analytic results for these parameters agree with the numerical
output from KEPLER within a factor of 2 in all nine models. The
agreement between the numerical and analytic results is the best for the
final entropy per baryon in the wind, as can be expected from the shallow
dependence on the correction factor $C$ in equations (48a) and (48b). This
correction factor indicates the uncertainties in the analytic treatment of
various neutrino heating processes. In our numerical models, neutrino
absorption on free nucleons is the main heating process. Neutrino-electron
scattering can make significant contribution to the heating for the cases
with high entropies, i.e., the models with $R=10$ km. For all nine models,
neutrino-antineutrino annihilation contributes at most $\sim 15\%$ to the
total heating.

While the general agreement between numerical and analytic results is important
to show, we also highlight the predicted power-law dependences on the neutrino
luminosity and mean energy, and the mass and radius of the neutron star for
the physical parameters listed in Table 1. As can be easily verified from
the numerical results in Table 1, the power-law dependences on the neutrino
luminosity for these physical parameters follow the analytic predictions
with good accuracy.

\subsection {Post-Newtonian calculations}

The models of the previous subsection were all calculated using a simple
Newtonian theory of gravity. A neutron star is of course a relativistic
system. To zeroth order the gravity of the neutron star can simply be increased,
e.g., by turning up its mass, as was done in Models 10D, 10E, and 10F.
However, the actual radial dependence of gravity differs from $r^{-2}$ in
general relativity. To check that this does not have an appreciable effect
on the mass outflow rate, entropy, and dynamic time scale, two additional models
were calculated using a version of KEPLER modified to include first-order
post-Newtonian corrections to the gravitational force (e.g., Fuller,
Woosley, \& Weaver 1986; Shapiro \& Teukolsky 1983). The gravitational
term in the force equation (Weaver, Zimmerman, \& Woosley 1978), $Gm_r/r^2$,
was replaced by
$${{Gm_r} \over {r^2}} \, \left(1 \, + \, {{P} \over {\rho c^2}} \, + \,
{{4 \pi P r^3} \over {m_r c^2}} \right) \left(1 \, - \, {{2 G m_r} \over {r
c^2}}\right)^{-1},$$
where $m_r$ is the mass included within a radius $r$.
The term corresponding to $\dot q$ in the energy equation was unchanged.

This includes the leading post-Newtonian corrections to the wind dynamics
(with the wind velocity always much less than $c$). Unfortunately it does not
include the gravitational effects on the neutrino energy spectra and 
angular distributions.

Models 10B and 10E were then recalculated using the 
modified gravitational field.
The former had a neutron star mass at infinity of 1.4
$M_{\sun}$ and a total neutrino luminosity of $6 \times 10^{51}$ erg s$^{-1}$. 
In steady state, the
post-Newtonian model had a mass loss rate of $7.0 \times 10^{-6}$
$M_{\sun}$ s$^{-1}$, an asymptotic entropy of 116, and a dynamic time scale 
of 0.044 s.
Compared to quantities in Table 1, we see that the effect of the modified
gravitational potential is to increase the entropy by about 1/3 (from 87
to 116) and decrease the dynamic time scale by about 1/3. The post-Newtonian 
version
of Model 10B in fact resembles Model 10E, though perhaps an effective mass
of 1.8 $M_{\sun}$ instead of 2 $M_{\sun}$ might have been more appropriate.

We also considered the extreme case of Model 10E (neutron star mass at
infinity of 2.0 $M_{\sun}$) {\sl and} post-Newtonian corrections. Neglecting
general
relativistic corrections to the neutrino transport is clearly wrong here.
While allowed by causality, this is also a heavier neutron star than
commonly derived in binary systems. We consider this as an example of 
the state a
neutron star might pass through while evolving to a black hole on a
Kelvin-Helmholtz cooling time scale. The asymptotic entropy for this model is
205, the mass loss rate is $2.4 \times 10^{-6}$ $M_{\sun}$ s$^{-1}$, and the
dynamic time
scale is 0.035 s; i.e., a 60\% increase in the entropy and about a factor of 2
reduction in the dynamic time scale.

In summary, the effect of post-Newtonian corrections to gravity is to
increase the entropy and shorten the dynamic time scale by substantial amounts.
In this regard, a stronger gravitational potential is more
favorable to the $r$-process.

\subsection {Effect of an external boundary pressure}

Except in the case of the accretion-induced collapse of a neutron star
the wind will not be able to flow unencumbered to infinity. It will be
impeded by external matter and radiation in that portion of the supernova
through which the shock has already passed.

Woosley \& Weaver (1995) show that the pressure behind the shock will be
largely due to radiation in a nearly isothermal sphere. At a time of several
seconds this temperature is very approximately two billion K. It declines
to under one billion K by the time the neutron star is 10 seconds old.

To illustrate the effect, we have recalculated Model 10E with an external
boundary pressure appropriate to a temperature of $2 \times 10^9$ K. 
This is an extreme
case; a realistic model would have less boundary pressure at 10 s. The
converged model was run for one second with this boundary pressure and
reached an approximate steady state though the outer boundary was being
gradually pushed out at 200 km s$^{-1}$ (corresponding to $PdV$ work of $4
\times 10^{46}$ erg s$^{-1}$). Figure 3 shows that there is little change in
the structure of the wind, except near its outer boundary. The asymptotic
entropy was increased from 129 to 140 and the mass loss rate reduced from
$6.7 \times 10^{-6}$ to about $6.3 \times 10^{-6}$ $M_{\sun}$ s$^{-1}$. The 
dynamic time
scale (at 0.5 MeV) was lengthened from 0.066 s to 0.11 s.

In summary, the effect of a moderate external boundary pressure is to
increase the entropy slightly, and decrease the expansion rate appreciably.
Overall there is little modification of the parameters of the wind in the
region where the $r$-process occurs except that the material has a longer
time to capture neutrons. The slower time scale at temperatures of 3 to 5
billion K will also give a smaller neutron-to-seed ratio in the ejecta.
This is an adverse effect, but not a large one.

\subsection {Effect of an artificial energy source}

As Figs. 1 and 2 show, the mass outflow rate is determined closer to the
neutron star than the entropy. One consequence of this is that any attempt
to increase the entropy of the wind simply by raising the overall neutrino
heating rate 
is thwarted by an increased outflow rate. The extra energy goes into doing
additional work against the gravity of the neutron star and not into raising 
the entropy of the wind. This is also to be expected from the dependences of
$S$ and $\dot M$ on the neutrino luminosity and the correction factor $C$
in equations (48a), (48b), (58a) and (58b): $S\propto(CL_\nu)^{-1/6}$ vs.
$\dot M\propto(CL_\nu)^{5/3}$.

Conversely, energy that is added after the mass loss rate has already been
determined can have a very beneficial effect, both in raising the entropy
and increasing the velocity, both good for the $r$-process. Energy 
that is added too late however, after the temperature has already gone below
two billion K, has little nucleosynthetic effect. The neutron-to-seed
ratio for the $r$-process has already been set.
This suggests that an additional energy source between about 20 and 50 km
might have considerable leverage on the nucleosynthesis in the wind.
Possible sources of this energy are discussed later in the paper ($\S6$).

We thus explored the effect of adding an additional 10$^{48}$ erg s$^{-1}$
to Model 10E at a radius between 20 and 30 km. The total energy deposited
thus became $4 \times 10^{48}$ erg s$^{-1}$ instead of $3 \times 10^{48}$
erg s$^{-1}$. Of this 10$^{48}$ erg s$^{-1}$, roughly 10\% has a physical
origin in the recombination of nucleons to $\alpha$-particles ($\sim7 \times
10^{18}$ erg g$^{-1}$ times $\dot M$) . This occurs at $T\lesssim 1$ MeV 
or a radius
of about 15--20 km.

The mass outflow rate in the modified model increased
slightly from $6.7 \times 10^{-6}$ $M_{\sun}$ s$^{-1}$ to $9.0 \times 10^{-6}$
$M_{\sun}$ s$^{-1}$, and the asymptotic entropy rose substantially, from 129 to
190. Most importantly the expansion time scale at 0.5 MeV decreased from
0.066 to 0.010 s. The velocity at 0.5 MeV was 3500 km s$^{-1}$. These are
conditions quite favorable to the $r$-process (Hoffman et al. 1996b).

Model 10F in Fig. 4 shows what a relatively small amount of energy will do
if it can be generated in the right place and time. An additional volume
energy term ($9.7 \times 10^{27}$ erg cm$^{-3}$ s$^{-1}$) was added for radii
between 15 and 25 km giving a total added energy of $5 \times 10^{47}$ erg
s$^{-1}$, a moderate perturbation on the $1.2 \times 10^{48}$ erg s$^{-1}$
the neutrinos were already adding. As a result, the mass loss rate increased
a little to $3.7 \times 10^{-6}$ $M_{\sun}$ s$^{-1}$, the entropy climbed to
192 and the dynamic time scale at 0.5 MeV shrank to 0.022 s. 
These conditions should
be quite favorable for the $r$-process for $Y_e \lesssim 0.40$ (Hoffman et
al. 1996b).

Larger energy deposition was not explored as it would no longer be a
perturbation on the neutrino-driven wind, but the driving term for the mass
loss.

\subsection {Uncertainty in the neutrino energy deposition}

Of the three interactions that deposit energy in the wind --- neutrino capture
on nucleons, neutrino scattering on electrons, and neutrino-antineutrino
annihilation ---
the last is probably worst approximated in our calculations. The radial
dependence of equation (14) is very steep. Detailed and better treatment
of this dependence was given by Janka (1991a,b).
Multidimensional and multi-group neutrino transport calculations by Jim
Wilson (1996) show that within two neutron star radii,
the approximation in equation (14) is probably a gross underestimate of 
the actual heating
rate due to neutrino-antineutrino
annihilation. Even larger deviations may be expected when a fully general
relativistic transport calculation is ultimately done. The effects of
general relativity on neutrino transport calculations were also 
discussed by Janka (1991a).

For someone seeking high entropy and rapid expansion neutrino-antineutrino
annihilation
has the attractive feature that it deposits energy at a rate independent of
the local matter density. It can thus boost the specific energy and
entropy of a wind after its density has already declined. The problem is the
steep fall-off of the neutrino-antineutrino annihilation rate with radius and 
its low efficiency.
For Model 10E, of the $3 \times 10^{48}$ erg s$^{-1}$ being deposited in the
steady-state model, about 10\% is due to neutrino-antineutrino annihilation.

We thus experimented with varying the radial dependence and efficiency of
neutrino-antineutrino annihilation. In one calculation Model 10E was run with an
additional 10$^{48}$ erg s$^{-1}$ deposited as a constant volume term ($3.4
\times 10^{28}$ erg cm$^{-3}$ s$^{-1}$) between 10 and 20 km (the surface
of the neutron star and twice its radius). As a result, the asymptotic
entropy increased from 129 to 137, the mass outflow rate climbed from $6.7
\times 10^{-6}$ to $8.8 \times 10^{-6}$ $M_{\sun}$ s$^{-1}$, and the dynamic
time scale
at 0.5 MeV declined from 0.066 s to 0.040 s.

A second calculation maintained the functional form of equation (14), 
but multiplied
the neutrino-antineutrino annihilation efficiency by a factor of six inside
20 km. This resulted in increases of the energy deposition to $3.4 \times
10^{48}$ erg s$^{-1}$ and mass outflow rate to $7.8\times10^{-6}\ M_{\sun}\ 
{\rm s}^{-1}$, and decreases in entropy to 127 and dynamic time scale to
0.061 s.

Finally, we explored, again in Model 10E, changing the radial dependence of
equation (14). The term $(1-x)$ in $\Phi(x)$ was raised to the third power 
instead
of the fourth and the overall efficiency multiplied by 3 (at all radii). This
raised the energy deposition to $3.8 \times 10^{48}$ erg s$^{-1}$, the
entropy to 134, the mass outflow rate to $8.8 \times 10^{-6}$ $M_{\sun}$ 
s$^{-1}$, and decreased the dynamic time scale to 0.044 s.

We conclude that reasonable variations in the efficiency of 
neutrino-antineutrino
annihilation can give a modest increase in the entropy and decrease in the 
dynamic time
scale. However, very large increases in the entropy, like a doubling, would
require more radical alterations than we think are justified at the
present time. Unfortunately any energy source that is peaked near the
neutron star surface tends to increase the mass loss rate more than 
the entropy.

\subsection {A time variable neutrino luminosity?}

Our analytic solutions assume a steady state in which the neutrino luminosity
does not vary. In the real situation the local neutrino flux in the
outflowing wind may vary considerably because of rotation, accretion, and
convective flows. We thus explored the effect on Model 10E of varying the
total neutrino luminosity on a time scale comparable to the flow time across
the wind zone. In particular, we take the total neutrino luminosity to be
$L_{\nu,\rm{tot}} = 6 \times 10^{51} [1 + 0.5 \sin
(2 \pi t/\tau_{\nu})]$ erg s$^{-1}$ with $\tau_{\nu}$ = 0.1 s. Much shorter
time scales would have had no effect. Much longer time scales would
have given the steady-state solutions previously described.

The chief effect of this variation in the neutrino luminosity was to cause 
a periodic variation in the
outflow velocity and dynamic time scale. The entropy, e.g., at 0.5 MeV, did not
change greatly from 129 in the steady-state model, being 127 $\pm$ 3
during three oscillations. But the outflow velocity varied almost linearly
with the neutrino luminosity so that the dynamic time scale at 0.5 MeV 
oscillated 
between
0.044 and 0.15 s (the steady-state value for Model 10E was 0.066 s). A
similar calculation in which $\tau_{\nu}$ = 0.2 s gave the same range of 
dynamic time
scales, but with an entropy oscillating between 115 and 136. In both cases
the larger entropy was associated with the faster outflow. Note that an
inverse dependence of the expansion time scale on the neutrino luminosity
is predicted by equation (61) for the steady-state wind.

We conclude that reasonable variations in the neutrino luminosity on 
intervals of
order 0.1 s can give a large range of expansion time scales at nearly
constant entropy. Thus it is possible to have, in a fraction of the ejecta,
material which has experienced the high entropy appropriate to a low 
time-averaged
neutrino luminosity, but with the rapid expansion time scale 
characteristic of the
temporary peaks.

\section{Electron fraction in the neutrino-driven wind}

In this section, we describe how the electron fraction $Y_e$ is
determined in the neutrino-driven wind. As we will see, the final
$Y_e$ in the ejecta is principally set by the characteristics of the $\nu_e$
and $\bar\nu_e$ fluxes. In this regard, we also discuss general aspects
of neutrino emission in supernovae, and present a novel neutrino ``two-color
plot" to illustrate the time evolution of the $\nu_e$ and $\bar\nu_e$ energy
distributions. To conclude the section, we discuss the implications of this
evolution plot for heavy element nucleosynthesis in supernovae.

\subsection{Input neutrino physics}

We begin our discussion by calculating the rates for the forward and reverse
reactions in equations (8a) and (8b). These reactions are the most important 
processes which set the electron fraction $Y_e$ in the neutrino-driven wind.
The cross sections for the forward reactions in equations
(8a) and (8b) are given by
\begin{equation}
\sigma_{\nu N}\approx {1+3\alpha^2\over \pi}G_F^2P_eE_e\approx
{1+3\alpha^2\over \pi}G_F^2E_e^2,
\end{equation}
where $\alpha\approx 1.26$, $P_e$ and $E_e$ are the momentum and total energy,
respectively, of the electron or positron in the final state, and where we 
have made the approximation $P_eE_e\approx E_e^2$. This approximation is
very good because (i) the energy of the final state
electron in equation (8a) is at least the neutron-proton mass difference; 
and (ii) in any
case, the typical neutrino energy is about 10 MeV or more, which makes the
final state electrons and positrons extremely relativistic.

Following the prescription for the neutrino flux in $\S3.1$, we find the rates
for the forward reactions in equations (8a) and (8b) to be
\begin{mathletters}
\begin{eqnarray}
\lambda_{\nu_en}&\approx&{1+3\alpha^2\over 2\pi^2}G_F^2{L_{\nu_e}\over R_\nu^2}
\left(\epsilon_{\nu_e}+2\Delta+{\Delta^2\over\langle E_{\nu_e}\rangle}\right)
(1-x),\\
\lambda_{\bar\nu_ep}&\approx&{1+3\alpha^2\over 2\pi^2}G_F^2{L_{\bar\nu_e}\over
R_\nu^2}\left(\epsilon_{\bar\nu_e}-2\Delta+{\Delta^2\over\langle E_{\bar\nu_e}
\rangle}\right)(1-x),
\end{eqnarray}
\end{mathletters}
where $\Delta=1.293$ MeV is the neutron-proton mass difference. At $r\gg 
R_\nu$, we can write the above rates as
\begin{mathletters}
\begin{eqnarray}
\lambda_{\nu_en}&\approx& 4.83L_{\nu_e,51}\left(\epsilon_{\nu_e,\rm MeV}+
2\Delta_{\rm MeV}+1.2{\Delta_{\rm MeV}^2\over \epsilon_{\nu_e,\rm MeV}}\right)
r_6^{-2}\ {\rm s}^{-1},\\
\lambda_{\bar\nu_ep}&\approx& 4.83L_{\bar\nu_e,51}\left(\epsilon_{\bar\nu_e,
\rm MeV}
-2\Delta_{\rm MeV}+1.2{\Delta_{\rm MeV}^2\over \epsilon_{\bar\nu_e,\rm MeV}}
\right) r_6^{-2}\ {\rm s}^{-1},
\end{eqnarray}
\end{mathletters}
where $\Delta_{\rm MeV}$ is $\Delta$ in MeV, and where we have taken an
average value of $\epsilon_\nu/\langle E_\nu\rangle=1.2$. In fact,
$\epsilon_\nu/\langle E_\nu\rangle$ ranges from 16/15 to 4/3 for neutrino 
energy distributions of the form $f(E_\nu)\propto E_\nu^2/[\exp(aE_\nu+b)+1]$
with $a>0$ and $-\infty<b<\infty$. We have taken into account the 
neutron-proton mass difference to give accurate values of the above reaction
rates. As we will see in $\S5.2$, the final $Y_e$ in the ejecta is sensitive 
to the ratio of these reaction rates.

The cross sections for the reverse reactions in equations (8a) and (8b) are
given by
\begin{equation}
\sigma_{eN}\approx{1+3\alpha^2\over 2\pi}G_F^2E_\nu^2,
\end{equation}
where $E_\nu$ is the neutrino energy in the final state. The rates for these
reactions are
\begin{equation}
\lambda_{e^-p}\approx\lambda_{e^+n}\approx 0.448T_{\rm MeV}^5\ {\rm s}^{-1},
\end{equation}
where we have neglected the neutron-proton mass difference, and have assumed
that the initial state electrons and positrons are extremely relativistic.
These approximations are reasonable at $T\gtrsim 1$ MeV, and become invalid
when $T$ approaches 0.5 MeV. However, these reaction rates are negligible
compared with those in equations (65a) and (65b) at $T\lesssim 1$ MeV.
This is because at these low temperatures, (i) the number density of
electron-positron pairs decreases significantly; and (ii) the cross sections
decrease rapidly, especially so for electron capture on proton, which has
to overcome the neutron-proton mass difference.
Therefore, the breaking-down of the above approximations has no serious
consequences for setting the final value of $Y_e$ in the ejecta.

\subsection{Determination of $Y_e$}
As noted in $\S2$, the electron fraction $Y_e$ in the ejecta is governed by
equation (7). We can rewrite this equation as
\begin{equation}
\dot Y_e=\lambda_1-\lambda_2 Y_e,
\end{equation}
where $\lambda_1=\lambda_{\nu_en}+\lambda_{e^+n}$, $\lambda_2=\lambda_1
+\lambda_{\bar\nu_ep}+\lambda_{e^-p}$, and $\dot Y_e=dY_e/dt=v(dY_e/dr)$.
Regardless of the particular forms of $\lambda_1$ and $\lambda_2$, the general
solution to the above equation is given by
\begin{equation}
Y_e(t)=\left[Y_e(0)-{\lambda_1(0)\over\lambda_2(0)}\right]I(0,t)+
{\lambda_1(t)\over\lambda_2(t)}-\int_0^tI(t^\prime,t){d\over dt^\prime}
\left[{\lambda_1(t^\prime)\over\lambda_2(t^\prime)}\right]dt^\prime,
\end{equation}
where $t=0$ is taken as the time when the ejecta leave the neutron star
surface at radius $R$, and where
\begin{equation}
I(t^\prime,t)=\exp\{-\int_{t^\prime}^t\lambda_2(t^{\prime\prime})
dt^{\prime\prime}\}
\end{equation}
is a ``memory'' function of the interaction history, with $\int_{t^\prime}^t
\lambda_2(t^{\prime\prime})dt^{\prime\prime}$ the total number of interactions
on a pair of neutron and proton between $t^\prime$ and $t$ (Qian 1993).

It is easy to show that the first term on the right-hand side of equation (69)
quickly vanishes at $t>0$. The gravitational binding energy of a nucleon at
the neutron star surface is $\sim GMm_N/R\sim 200$ MeV. A nucleon has to
obtain at least this amount of energy from the neutrino fluxes in order to
escape to large radii. The main heating reactions, $\nu_e$ and $\bar\nu_e$
absorption on free nucleons, are also responsible for determining $Y_e$.

From each interaction with the $\nu_e$ or $\bar\nu_e$ flux, an amount of
energy $\sim 10$--20 MeV per nucleon is absorbed. 
Therefore, a nucleon in the ejecta has to have at
least 10 interactions above the neutron star surface, and the total number
of interactions on a pair of neutron and proton in the ejecta
satisfies
\begin{equation}
\int_0^\infty\lambda_2(t)dt>\int_0^\infty[\lambda_{\nu_en}(t)
+\lambda_{\bar\nu_ep}(t)]dt>20,
\end{equation}
where $t=\infty$ is a symbolic time when $\lambda_2$ becomes negligibly small.

From the discussions in $\S3.3$, we know that most of the heating interactions
take place
at temperatures around $T_{\rm eff}\sim 2$ MeV. So for
$T\lesssim 2$ MeV, we can safely assume $I(0,t)\ll 1$ and neglect the first 
term on the right-hand side
of equation (69).

Making use of the relation
\begin{equation}
{d\over dt^\prime}I(t^\prime,t)=\lambda_2(t^\prime)I(t^\prime,t),
\end{equation}
we can perform integration by parts in the third term on the right-hand side
of equation (69), and obtain
\begin{equation}
Y_e(t)\approx{\lambda_1(t)\over\lambda_2(t)}-{1\over\lambda_2(t)}{d\over dt}
\left[{\lambda_1(t)\over\lambda_2(t)}\right]
\end{equation}
when $I(0,t)\ll 1$.
In deriving the above equation, we have assumed
\begin{equation}
{1\over\lambda_2(t)}\left|{d\over dt}\left[{\lambda_1(t)\over\lambda_2(t)}
\right]\right|
\ll{\lambda_1(t)\over\lambda_2(t)},
\end{equation}
and therefore, have neglected terms of higher orders in $\lambda_2^{-1}(t)$,
i.e., the remaining integral
$$\int_0^tI(t^\prime,t){d\over dt^\prime}\left\{{1\over\lambda_2(t^\prime)}
{d\over dt^\prime}\left[{\lambda_1(t^\prime)\over\lambda_2(t^\prime)}\right]
\right\}dt^\prime.$$

The condition in equation (74) is called the quasi-equilibrium (QSE) condition.
This is because under this condition,
\begin{equation}
Y_e(t)\approx Y_{e,\rm EQ}(t)-{1\over\lambda_2(t)}{d\over dt}Y_{e,\rm EQ}(t)
\approx Y_{e,\rm EQ}(t),
\end{equation}
where
\begin{equation}
Y_{e,\rm EQ}\equiv {\lambda_1(t)\over\lambda_2(t)}
\end{equation}
is the instantaneous equilibrium value at time $t$. From the rates and
discussions in $\S5.1$, we see that for $T\lesssim 1$ MeV, $\lambda_1$ and
$\lambda_2$ are dominated by the rate(s) for neutrino absorption on free
nucleons. Consequently, $d(\lambda_1/\lambda_2)/dt\approx 0$, and $Y_e$
assumes a constant equilibrium value
\begin{equation}
Y_{e,f}\approx {\lambda_{\nu_en}\over\lambda_{\nu_en}+\lambda_{\bar\nu_ep}}
\approx \left(1+{L_{\bar\nu_e}\over L_{\nu_e}}{\epsilon_{\bar\nu_e}-2\Delta
+1.2\Delta^2/\epsilon_{\bar\nu_e}\over 
\epsilon_{\nu_e}+2\Delta+1.2\Delta^2/
\epsilon_{\nu_e}}\right)^{-1}
\end{equation}
for $T\lesssim 1$ MeV (Qian et al. 1993).

The $Y_{e,f}$ given by equation (77) would be the final value of $Y_e$ in
the ejecta if equations (7) and (68) were valid for all temperatures. However,
as we have discussed before, free nucleons begin to be bound into 
$\alpha$-particles
and heavier nuclei at $T<1$ MeV (cf. eq. [62]). Therefore, 
equations (7) and (68)
are to be replaced by
\begin{equation}
\dot Y_e=\lambda_1^\prime
-\lambda_2^\prime Y_e
\end{equation}
at $T<1$ MeV, where $\lambda_1^\prime=\lambda_{\nu_en}+
(\lambda_{\bar\nu_ep}-\lambda_{\nu_en})(1-X_N)/2$ and $\lambda_2^\prime
=\lambda_{\nu_en}
+\lambda_{\bar\nu_ep}$. In deriving equation (78), 
we have neglected the rates for
neutrino absorption on $\alpha$-particles and heavier nuclei and all the
electron capture rates, compared with the neutrino absorption rates on
free nucleons. We have also assumed $Y_e\approx X_p+(1-X_N)/2$, where
$X_N=X_p+X_n$ is the mass fraction of free nucleons, with $X_p$ and $X_n$
the mass fractions of protons and neutrons, respectively. This approximation
for $Y_e$ is good for material mainly composed of free nucleons and
$\alpha$-particles.

Equation (78) is mathematically identical to equation (68). 
At the transition from equation (68) to (78), we have
$Y_e\approx \lambda_1/\lambda_2\approx \lambda_1^\prime/\lambda_2^\prime$
and $d(\lambda_1/\lambda_2)/dt\approx d(\lambda_1^\prime/\lambda_2^\prime)/dt
\approx 0$. From the general solution given in equation
(69), we can easily show that
\begin{equation}
Y_e(t)\approx{\lambda_1^\prime(t)\over\lambda_2^\prime(t)},
\end{equation}
provided the QSE condition
\begin{equation}
{1\over \lambda_2^\prime(t)}\left|{d\over dt}\left[{\lambda_1^\prime(t)\over
\lambda_2^\prime(t)}\right]\right|\ll{\lambda_1^\prime(t)
\over\lambda_2^\prime(t)}
\end{equation}
is satisfied. From the expression for $\lambda_1^\prime$, we see that the 
QSE condition in equation (80) depends on how fast $X_N$ is changing. In
turn, the rate at which $X_N$ changes is determined by the thermodynamic
and hydrodynamic conditions in the ejecta (cf. eq. [62]). 

Typically, the
QSE condition in equation (80) is no longer met when
\begin{equation}
\lambda_2^\prime(r){r\over v(r)}\sim 1.
\end{equation}
At this point, the neutrino absorption reactions are not frequent enough
to change $Y_e$ any more, and $Y_e$ ``freezes'' out at the last achievable
instantaneous equilibrium value
\begin{equation}
Y_{e,f}^\prime\approx{\lambda_{\nu_en}\over \lambda_{\nu_en}+
\lambda_{\bar\nu_ep}}+{1-X_N\over 2}{\lambda_{\bar\nu_ep}-\lambda_{\nu_en}\over
\lambda_{\nu_en}+\lambda_{\bar\nu_ep}}.
\end{equation}
Therefore, if the QSE condition for equation (78) is still satisfied when a 
significant
fraction of the material is in $\alpha$-particles and heavier nuclei, the
final $Y_e$ in the ejecta will deviate from $Y_{e,f}$ given in equation (77).
The influence of $\alpha$-particles and heavier nuclei on $Y_e$ in the ejecta
was first pointed out by Fuller \& Meyer (1995). 

Because the evolution of
$Y_e$ in the ejecta at $T<1$ MeV is coupled with the change in the
nuclear composition of 
the ejecta, and depends sensitively on the thermodynamic and hydrodynamic
conditions in the ejecta (cf. eqs. [62] and [81]), we will follow $Y_e$ 
numerically in the
nucleosynthesis calculations to be presented in a separate paper (Hoffman
et al. 1996b).
However, whether the final $Y_e$ in the ejecta is given by equation (77) or
(82), mathematically, $Y_e=0.5$ corresponds to
\begin{equation}
\lambda_{\bar\nu_ep}\approx\lambda_{\nu_en},
\end{equation}
barring the unlikely case where the QSE condition for equation (78) holds
until all the free nucleons disappear. The qualitative feature of heavy
element nucleosynthesis depends crucially on whether $Y_e>0.5$ or $Y_e<0.5$.
From equations (65a), (65b), and (83), we see that neutrino luminosities
and energy distributions have direct bearings on the nature of heavy element
nucleosynthesis in supernovae. 

\subsection{Neutrino two-color plot}

As the neutrinos diffuse towards the edge of the neutron star, their
interactions with matter decrease with the decreasing temperature and
density. Gradually, they decouple from the thermodynamic equilibrium with the
neutron star matter, keeping energy distributions corresponding to the
conditions (e.g., temperature, density, chemical composition) in the
decoupling region. Because only energy-exchange processes are directly
responsible for achieving thermodynamic equilibrium, and different neutrino
species have different ability to exchange energy with matter, the
thermodynamic decoupling processes of different neutrino species occur in
different regions of the neutron star. In turn, each neutrino species emerges
from the neutron star with its own characteristic energy distribution.

All neutrino species exchange energy with electrons. In fact, the
neutral-current scattering on electrons is the main energy-exchange process
for $\nu_\mu$, $\bar\nu_\mu$, $\nu_\tau$, and $\bar\nu_\tau$. On the other
hand, although $\nu_e$ and $\bar\nu_e$ have the extra charged-current
scattering on electrons which $\nu_\mu$, $\bar\nu_\mu$, $\nu_\tau$, and
$\bar\nu_\tau$ lack, they exchange energy with matter even more efficiently
through the charged-current absorption reactions on nucleons given in
equations (8a) and (8b). Therefore, $\nu_\mu$, $\bar\nu_\mu$, $\nu_\tau$, and
$\bar\nu_\tau$ decouple at higher temperature and density inside the neutron
star than $\nu_e$ and $\bar\nu_e$, and correspondingly, their energy
distributions are harder than those of $\nu_e$ and $\bar\nu_e$.

As we have seen in $\S5.2$, the luminosities and energy distributions for
$\nu_e$ and $\bar\nu_e$ have significant effects on $Y_e$, or equivalently, 
the neutron excess in the
wind. In addition, emission of $\nu_e$ and $\bar\nu_e$ is coupled with the
deleptonization process of the neutron star, which makes the problem of
$\nu_e$ and $\bar\nu_e$ transport in the neutron star even more complicated.
In what follows, we give a more detailed discussion of the time
evolution of $\nu_e$ and $\bar\nu_e$ emission in supernovae.

Shortly after the creation of the neutron star by the core collapse and
subsequent supernova explosion, its neutron concentration is larger than,
but still comparable to its proton concentration. Because $\nu_e$ and
$\bar\nu_e$ exchange energy with matter mainly through absorption on
neutrons and protons, respectively, $\nu_e$ can stay in thermodynamic
equilibrium with the neutron star matter down to somewhat lower temperatures
and densities than $\bar\nu_e$. In other words, $\bar\nu_e$ decouple at
slightly higher temperatures than $\nu_e$, and have a slightly larger
average energy, i.e., $\langle E_{\bar\nu_e}\rangle\gtrsim \langle E_{\nu_e}
\rangle$. The luminosities for $\nu_e$ and $\bar\nu_e$ are about the same,
so the number flux of $\nu_e$ is larger than that of $\bar\nu_e$. This results
in a net electron lepton number flux leaking out of the neutron star, which
reflects the net deleptonization process of converting protons into neutrons
through electron capture inside the neutron star. Meanwhile, the neutron
star is shrinking to its final equilibrium configuration. The heating
caused by the compression raises the temperature of the neutron star interior.
As a result, both $\langle E_{\bar\nu_e}\rangle$ and $\langle E_{\nu_e}\rangle$
increase initially while keeping $\langle E_{\bar\nu_e}\rangle \gtrsim\langle
E_{\nu_e}\rangle$.

Because of the diffusive nature of the neutrino transport processes, we expect
to see the strongest deleptonization in the surface layers of the neutron star.
As these surface layers become more and more neutron rich due to the
deleptonization, the decoupling regions for $\nu_e$ and $\bar\nu_e$ are
progressively separated from each other. The deleptonization effects on
$\langle E_{\nu_e}\rangle$ and $\langle E_{\bar\nu_e}\rangle$ are most
pronounced when the neutron star approximately reaches its final radius and
the compressional heating becomes negligible. At this point, $\langle 
E_{\bar\nu_e}\rangle$ still increases because $\bar\nu_e$ decouple deeper
with the deleptonization. On the contrary, $\langle E_{\nu_e}\rangle$ begins
to decrease as $\nu_e$ decouple further out in the neutron star due to the
increasing neutron concentration in the surface layers.

Schematically, we can illustrate the time evolution of the $\nu_e$ and
$\bar\nu_e$ energy distributions on a two-dimensional plot. We choose the
abscissa and ordinate to be $\epsilon_{\nu_e}$ and $\epsilon_{\bar\nu_e}$,
respectively, because these energies are more directly related to our 
discussion of $Y_e$ in the ejecta (cf. eqs. [65a] and [65b]). Borrowing a
term from optical photometry, we refer to this plot as the ``neutrino
two-color plot.'' For a particular supernova model, the time evolution of
$\epsilon_{\nu_e}$ and $\epsilon_{\bar\nu_e}$ is represented by a contour
line on this plot. Taking $\epsilon_{\nu_e}$ and $\epsilon_{\bar\nu_e}$
as functions of time from Fig. 3 in Woosley et al. (1994), we present an
example plot in Fig. 5. The solid contour line in Fig. 5 shows the evolution
track of $\epsilon_{\nu_e}$ and $\epsilon_{\bar\nu_e}$. The open circles
on this track indicate the time after the onset of the collapse in intervals
of approximately 1/3 s for $t\approx0$--4 s and approximately 1 s for 
$t\approx4$--18 s.
Time $t\approx0$ corresponds to the open circle at the lower left corner of the
plot, and time increases thereafter along the track to $t\approx 18$ s 
at the last
open circle. The evolution of $\epsilon_{\nu_e}$ and $\epsilon_{\bar\nu_e}$
slows down considerably at $t>10$ s. As a result, there are two overlapping
open circles at $t\approx12$, 14, and 17 s (i.e., the open circles for 
$t\approx13$, 15, and 18 s cannot be distinguished). From this plot, 
we can see that $\epsilon_{\nu_e}$ and $\epsilon_
{\bar\nu_e}$ both increase for $t\approx0$--1 s, and $\epsilon_{\nu_e}$
starts to decrease while $\epsilon_{\bar\nu_e}$ increases for $t\gtrsim2$ s,
in agreement with our previous discussion. The evolution of $\epsilon_{\nu_e}$
and $\epsilon_{\bar\nu_e}$ during $t\approx1$--2
s is more complicated and may only be understood by more involved argument
than our previous qualitative discussion.

Numerical supernova calculations show that shortly after the explosion,
the ratio $L_{\bar\nu_e}/L_{\nu_e}$ always stays
close to 1 (see, e.g., Fig. 2 in Woosley et al. 1994). Therefore, 
without specific detailed neutrino
transport calculations, we can divide the neutrino two-color plot into three
regions corresponding to $Y_e<0.5$, $Y_e\approx 0.5$, and $Y_e>0.5$ in the
ejecta. The region corresponding to $Y_e\approx 0.5$ is bounded by the range
of values for $L_{\bar\nu_e}/L_{\nu_e}$, and is determined by equation (83).
In Fig. 5, we give an example situation where this region is bounded by
$L_{\bar\nu_e}/L_{\nu_e}\approx 1.1$ from below and by $L_{\bar\nu_e}/L_{\nu_e}
\approx 1.0$ from above. The parameter space to the left of this region
corresponds to $Y_e<0.5$, and the parameter space for $Y_e>0.5$ lies to the
right of this region. In general, our neutrino two-color plot can be presented
similarly for any supernova model. The implications of this plot for heavy
element nucleosynthesis in supernovae are discussed in the next subsection.

\subsection{Nucleosynthesis implications of the neutrino two-color plot}

As discussed in $\S1$, a severe failure of $r$-process calculations in the
neutrino-driven wind is the overproduction of nuclei in the vicinity of
$N=50$ at $t\sim 1$ s after the supernova explosion. Hoffman et al. (1996a)
have shown that this failure can be turned into an attractive feature if
$Y_e$ in the ejecta is slightly increased from the value in Wilson's
supernova model to $0.484\lesssim Y_e\lesssim 0.488$, for which some light
$p$-process nuclei can be produced. From Fig. 5, we see that the evolution
trajectory of $\epsilon_{\bar\nu_e}$ and $\epsilon_{\nu_e}$ for $t\sim 1$ s
lies entirely
in the region of $Y_e\approx 0.5$ on the neutrino two-color plot. This implies
that some small uncertainties in the supernova model predictions for 
$\epsilon_{\bar\nu_e}$, $\epsilon_{\nu_e}$, and/or $L_{\bar\nu_e}$, $L_{\nu_e}$
may explain the overproduction of the nuclei in the vicinity of $N=50$. It also
hints that future supernova models with more accurate neutrino transport
calculations may not encounter the overproduction problem, but would facilitate
the nucleosynthesis of some light $p$-process nuclei instead.

On the other hand, we see that $\epsilon_{\bar\nu_e}$ and $\epsilon_{\nu_e}$
predicted by Wilson's supernova model will clearly give $Y_e<0.5$ in the
neutrino-heated ejecta for $t>3$ s. For any $r$-process nucleosynthesis to
take place in the ejecta, we must have $Y_e<0.5$. Therefore, the neutrino
two-color plot in Fig. 5 at least signifies the possibility of an $r$-process
for $t>3$ s in Wilson's supernova model, even in the presence of some small
uncertainties in the predicted neutrino characteristics. 
While the exact values of $Y_e$ in the ejecta depend on the accuracy of
supernova neutrino transport calculations on the one hand, and the
thermodynamic and hydrodynamic conditions in the ejecta on the other hand, we
feel that the neutrino two-color plot has some interesting implications for
the nature of supernova nucleosynthesis. Perhaps the nucleosynthesis of some
light $p$-process nuclei can occur at $t\sim 1$ s, and the $r$-process
nucleosynthesis is produced later when the neutrino characteristics evolve to
the neutron-rich region of the neutrino two-color plot.

\section{Conclusions and Discussion}

The principal conclusions of our paper are simple analytic expressions,
verified by comparison to numerical simulation,
for the entropy (eqs. [48a]
and [48b]),
mass loss rate (eqs. [58a] and [58b]), and dynamic time scale (eq. [61]) in
the neutrino-driven wind. We have also given a detailed  
analysis of how $Y_e$ is determined ($\S5$) in the
wind.

We began this work in the hope that a simple physical model would yield the
conditions required by Woosley et al. (1994) for the $r$-process, in
particular rapid expansion time scales and entropies of order 300 or more.
While pursuing this quest, a great deal was learned about the properties of
neutrino-driven winds, which eventually came to appear quite simple. Mutually
confirming analytic and numerical calculations give the solutions in Table
1.  These models span a range of gravitational potential and neutrino
luminosity that should typify the Kelvin-Helmholtz evolution of most
neutron stars.

Unfortunately, none of the models in Table 1 will give a good $r$-process
unless $Y_e$ is very low (Hoffman et al. 1996b).  Nor can one continually
increase the entropy by turning the neutrino luminosity down.  The entropy
only increases as the sixth root of $L_{\nu}$ while the mass loss rate
declines as $L_{\nu}^{5/3}$ (assuming that the neutrino energy distributions
stay more or less the same over the period of interest, cf. Fig. 5). 
Moreover, the dynamic time scale
becomes unacceptably long. We conclude that one of the following must be
true: (1) the (heavy) $r$-process isotopes are not made in the 
neutrino-powered
wind of young neutron stars (of the type given in Table 1); (2) the
nuclear physics of the $r$-process is at fault (presently unlikely); or (3)
important physics has been left out of the models in Table 1. We shall
spend the remainder of the paper discussing the third possibility.

What has been left out? First, there are effects that we know exist and can
estimate --- general relativity and nucleon recombination. In $\S$4.2 we
provided numerical calculations in a post-Newtonian approximation to show
the effects of general relativity. Any effect that strengthens the gravitational
potential leads to higher entropy in the wind ($\S$3.3.1). The essential
consequences of general relativity on our models can be obtained simply by
increasing the effective mass of the neutron star --- e.g., Models 10DEF vs.
Models 10ABC. We also considered, however, in a post-Newtonian calculation,
the extreme case of a neutron star having a radius of 10 km and a
gravitational mass (at infinity) of 2 $M_{\sun}$.  This was the only calculation
without an artificial energy source that gave an entropy over 200. Perhaps
the heavy $r$-process is made in neutron stars that are on their way to
becoming black holes. However, one cannot extend this approach indefinitely.
At some point, not much beyond the extreme case considered, as the
gravitational potential deepens, the neutron star becomes unstable for any
equation of state and turns into a black hole on a dynamic (collapse) time
scale. Whether sufficient matter is ejected from such objects to make the
solar $r$-process abundances remains to be determined.

The recombination of nucleons into $\alpha$-particles releases 
an amount of energy of about $\Delta Q\sim7$ MeV
per nucleon. Assuming most of the nucleons recombine at $T\sim 0.5$ MeV,
we get an additional entropy
of about $\Delta S\sim \Delta Q/T\sim 14$.
This was not included in either the numerical or
analytic models and should be added to all the entropies calculated
in the paper.

Our treatment of neutrino-antineutrino annihilation is very 
approximate (cf. Janka 1991a,b) and there are
expectations that a more realistic treatment might lead to a moderate
increase in the entropy ($\S$4.5) and reduction in the dynamic time scale. Our
opinion is that this is too small a change to restore the $r$-process. 

It is rather unfortunate that we did not get or understand the high
entropies obtained in Wilson's supernova model used by Woosley et al. (1994)
for the $r$-process calculations. We note that Wilson's supernova model
derives from a fully general relativistic hydrodynamic code with a consistent
and detailed treatment of neutrino transport. However, the origin of the
discrepancies in entropy between our study and Wilson's numerical calculations
remains unknown and is still under investigation.

In $\S$4.6 we considered the effect of a time variable luminosity and found
that, while the entropy did not change much, considerable leverage could be
exercised on the expansion time scale.  Perhaps the $r$-process is
made in those regions of the wind that begin their ejection at low neutrino
luminosity, but absorb additional energy and expand faster as the local
neutrino luminosity increases.

Indeed, the dynamic time scale is an important third parameter not sufficiently
emphasized in previous studies of the $r$-process in the wind. 
For a given $Y_e$,
increasing the neutron-to-seed ratio can be achieved either by reducing the
reaction rates that assemble $\alpha$-particles into heavy nuclei, i.e., by
reducing the density (raising the entropy) or by decreasing the time during
which these reactions can operate. Since the rate of $\alpha$-burning goes
approximately as $\rho^3$ (Woosley \& Hoffman 1992), one can obtain
approximately the same $r$-process at one-half the requisite entropy if the
time scale is about 8 times shorter.  One general result of our study is
that it is easier to achieve a more vigorous outflow than it is to increase
the entropy, thus to some extent the quest for $S \gtrsim$ 300 may be
misguided. Entropies of $\sim 150$, which are readily achievable may 
function equally well
if the expansion time scale can be cut by an order of magnitude to $\sim$0.01 s.

The most effective way of accelerating the expansion, and, as it turns out,
of increasing the entropy, is to provide an additional source of energy of
order 10$^{48}$ erg s$^{-1}$ at between 1.5 and 3 times the neutron star
radius.  This is a region where the mass loss rate has, for the most part,
already been determined, but the freeze-out of $\alpha$-particles from NSE
($T \approx 5 \times 10^9$ K) has not yet occurred. The effect of a purely
arbitrary energy source was explored in $\S$4.4.  The origin of such
energies brings us to effects that might be there, but which are very
difficult to estimate.

First, we know that fall-back will occur as the supernova explosion develops.
Convection becomes less important to the explosion mechanism as time
passes, but material is still decelerated to below the escape speed as the
shock interacts with the stellar mantle. Woosley \& Weaver (1995) find that
typically $\sim$0.1 $M_{\sun}$ falls back in about 10$^4$ s (more earlier than
later), so one expects this accreted material to be adding mass faster than
the wind is removing it. However, the hydrodynamic interaction is
complex. It is uncertain whether {\sl any} material finds its way back to
the neutron star during the first 20 s of interest. But the entropy of the
infalling material is much lower and it may descend in plumes while the
wind rises in large bubbles. It may even be that the wind is derived from
the accreted material rather than the neutron star itself. It is difficult
to guess what the actual interaction will be of the plumes with the neutron
star, but it may be that the wind starts with an initial entropy higher
(from the accretion shock) than the steady-state value derived
here. Electron capture near the neutron star would also influence the 
starting value of $Y_e$.

A ten-second-old neutron star will also be the site of continuing energetic
activity other than its neutrino emission. In particular, the star may be
violently vibrating, rotating rapidly, or have a strong magnetic field.
Small oscillations at the base of the developing neutron star crust
will steepen into shocks in the steep density gradient at the
neutron star surface (Woosley 1996).
This must certainly occur at some level, but whether it leads to $\sim
10^{48}$ erg s$^{-1}$ being deposited at greater than 1.5 neutron star
radii for a ten-second-old neutron star is unclear. 

The magnetic field configuration near the surface of a newly born neutron
star is very uncertain --- both its spatial structure and strength --- but
there are speculations (Duncan \& Thompson 1992) that the magnetic field 
may be very large,
$\sim10^{15}$ gauss. Since the outer layers of the neutron star have recently
been convective, the field structure near the surface may be highly
tangled. A 10$^{15}$ gauss field could confine, or at least impede the
escape of plasma having a temperature of 4 MeV, i.e., greater than the
neutrinosphere temperature at late times. Such a field could have several
effects. To the extent that it impedes the outflow of the wind, its effect
resembles that of a stronger gravitational potential. Higher entropy is a
likely outcome. To the extent the field reconnects at higher altitudes,
because of rotational shearing or stretching in the wind, it provides the
desired extra heating. Flow in the field may also develop clumps, the
analogue of ``photon bubbles" studied in x-ray pulsars (Klein \& Arons
1989, 1991).

If the dynamic time scale and entropy are inadequate, one may have to turn to a
significant lowering of $Y_e$. As discussed in $\S5$, $Y_e$ in the ejecta
is determined by neutrino emission characteristics of the neutron star. In
turn, the neutrino luminosity and energy spectrum are determined by neutrino
interactions in the hot dense neutron star ($\S5.3$). Any uncertainty in
our understanding of neutrino interactions in hot dense matter causes an
inaccurate prediction of the emission properties. For example, some recent
studies (Keil, Janka, \& Raffelt 1995; Sawyer 1995) suggest that the
strength of neutrino interactions is reduced by the medium response of
nuclear matter, with respect to the case of neutrino interactions with a
single nucleon. Although it is not clear whether the effects found by these
studies on neutrino characteristics can change $Y_e$ in the ejecta
significantly, a complete picture of supernova nucleosynthesis
may await better knowledge of neutrino interactions in hot dense matter.

Finally, the neutrino characteristics above the neutron star surface can be
altered by the process of neutrino flavor transformation. For example, if
$\nu_\mu$ or $\nu_\tau$ has a cosmologically significant mass in the range of
1--100 eV and the mass of $\nu_e$ is much smaller, then $\nu_\mu$ or
$\nu_\tau$ will be transformed into $\nu_e$ above the neutron star surface
through the Mikheyev-Smirnov-Wolfenstein (MSW) mechanism (Fuller et al. 1992;
Qian et al. 1993). Because $\nu_\mu$ and $\nu_\tau$ have a higher average
energy than $\bar\nu_e$, conversion of $\nu_\mu$ or $\nu_\tau$ into $\nu_e$
will increase $Y_e$ in the ejecta (cf. eq. [77]). Such flavor transformation
is undesirable (Qian et al. 1993). On the other hand, if $\nu_e$ has a mass
of a few eV and $\nu_\mu$ or $\nu_\tau$ is much lighter, then $\bar\nu_\mu$
or $\bar\nu_\tau$ will be transformed into $\bar\nu_e$ above the neutron star
surface (Qian 1993; Qian \& Fuller 1995). In this case, 
conversion of $\bar\nu_\mu$ or
$\bar\nu_\tau$ into $\bar\nu_e$ can significantly reduce $Y_e$ in the ejecta
(Fuller, Qian, \& Wilson 1996). In view of the recent claim of possible
evidence for $\bar\nu_\mu\rightarrow\bar\nu_e$ oscillations by the LSND
experiment (Athanassopoulos et al. 1995) and other astrophysical evidence for
massive neutrinos (see, e.g., Fuller, Primack, \& Qian 1995), this latter 
process of
flavor transformation is worth further exploring.

\acknowledgments 
We are grateful to Rob Hoffman for making the first
version of the neutrino two-color plot for us, and for providing us with
the results of nuclear reaction network calculations for various
conditions. We thank the referee (H.-Thomas Janka) and Adam Burrows for
helpful comments on improving the paper.
We also want to thank Wick Haxton, Craig Hogan, Jim Wilson, and
especially George Fuller for interesting and informative conversations on
the subject of the paper. Y.-Z. Qian thanks Sterl
Phinney and Kip Thorne for reading the draft of the paper and Caltech for a
fellowship during the final stage of this work. This work was supported by
the Department of Energy under Grant No.  DE-FG06-90ER40561 at the
Institute for Nuclear Theory, where Y.-Z. Qian was a postdoctoral research
associate. Woosley was supported by NSF Grants No. AST 91-15367 
and No. AST 94-17161 at UCSC, and
by an Alexander von Humboldt Stiftung in Germany.
\vfill
\eject

\vfill\eject
\centerline {Figure Captions:}

{\bf Fig. 1} Conditions in the numerical model for a neutrino-driven wind
assuming Newtonian physics, a total neutrino luminosity in all flavors of $6
\times 10^{51}$ erg s$^{-1}$, and a neutron star mass and radius of 2.0
$M_{\sun}$ and 10 km, respectively (Model 10E). The top frame gives the
dimensionless entropy per baryon $S_{\rm tot}$ (long dashed line), rate for
net 
energy deposition by neutrinos, $\dot q$, in 10$^{20}$ erg g$^{-1}$ s$^{-1}$
(short dashed line), and mass outflow rate, $\dot M=4 \pi r^2
\rho v$, in units of 10$^{-5}$ $M_{\sun}$ s$^{-1}$ (solid line),
as functions of radius in km. Note the abrupt decline in
$\dot q$ at 35 km as the nucleons recombine. The bottom frame gives
the density, $\rho$, in 10$^{5}$ g cm$^{-3}$ (solid line), temperature, $T$,
in 10$^9$ K
(long dashed line), and outflow velocity, $v$, in 10$^7$ cm s$^{-1}$ 
(short dashed line),
all on the same radial scale.

\vskip 0.3 in

{\bf Fig. 2} Conditions in the numerical model for a neutrino-driven wind
from a neutron
star of 1.4 $M_{\sun}$ with a radius of 30 km and a total neutrino luminosity 
of $6
\times 10^{52}$ erg s$^{-1}$ (Model 30B). Notation and quantities 
edited are the same as
in Fig. 1, except that the density is in 10$^7$ g cm$^{-3}$ and the mass
outflow rate is in units of 10$^{-4}$ $M_{\sun}$ s$^{-1}$ for display
purposes.

{\bf Fig. 3} Effect of a boundary pressure equivalent to a boundary temperature
of $2\times10^9$ K in Model 10E. The total entropy, temperature, and 
outflow velocity are given as functions of radius. The solid lines are for
Model 10E without the boundary pressure, and the corresponding conditions
for the same model with the boundary pressure are shown as dashed lines.

{\bf Fig. 4} Conditions in the numerical model for a neutrino-driven wind
assuming Newtonian physics, a total neutrino luminosity of $3.6\times10^{51}$
erg s$^{-1}$, and a neutron star mass and radius of 2.0 $M_{\sun}$ and 10 km,
respectively (Model 10F),
when an
additional (artificial) energy source of $5 \times 10^{47}$ erg s$^{-1}$ is
evenly spread (by volume) between 15 and 25 km (1.5 and 2.5 neutron
star radii). Notation and quantities edited are the same as in Fig. 1.

{\bf Fig. 5} Neutrino two-color plot. The time evolution of the $\bar\nu_e$
and $\nu_e$ mean energies, $\epsilon_{\bar\nu_e}$ and $\epsilon_{\nu_e}$,
respectively, predicted by Wilson's supernova model,
is shown as a solid contour on the plot. The open circles on this contour 
indicate the time after the onset of the collapse from $t\approx 0$ at the
lower end to $t\approx 18$ s at the upper end. Time increases along the
contour in intervals of approximately 1/3 s for $t\approx0$--4 s and
approximately 1 s for $t\approx4$--18 s. Note that due to the slow
evolution of $\epsilon_{\bar\nu_e}$ and $\epsilon_{\nu_e}$ at $t>10$ s,
there are two overlapping open circles at $t\approx12$, 14, and 17 s (i.e.,
the open circles for $t\approx13$, 15, and 18 s cannot be distinguished).
Three regions, separated by
the two dashed lines, correspond to
values of $\epsilon_{\bar\nu_e}$ and $\epsilon_{\nu_e}$ which would give
$Y_e>0.5$, $Y_e\approx0.5$, and $Y_e<0.5$, respectively,
in the neutrino-driven wind.
See text for detailed explanation.

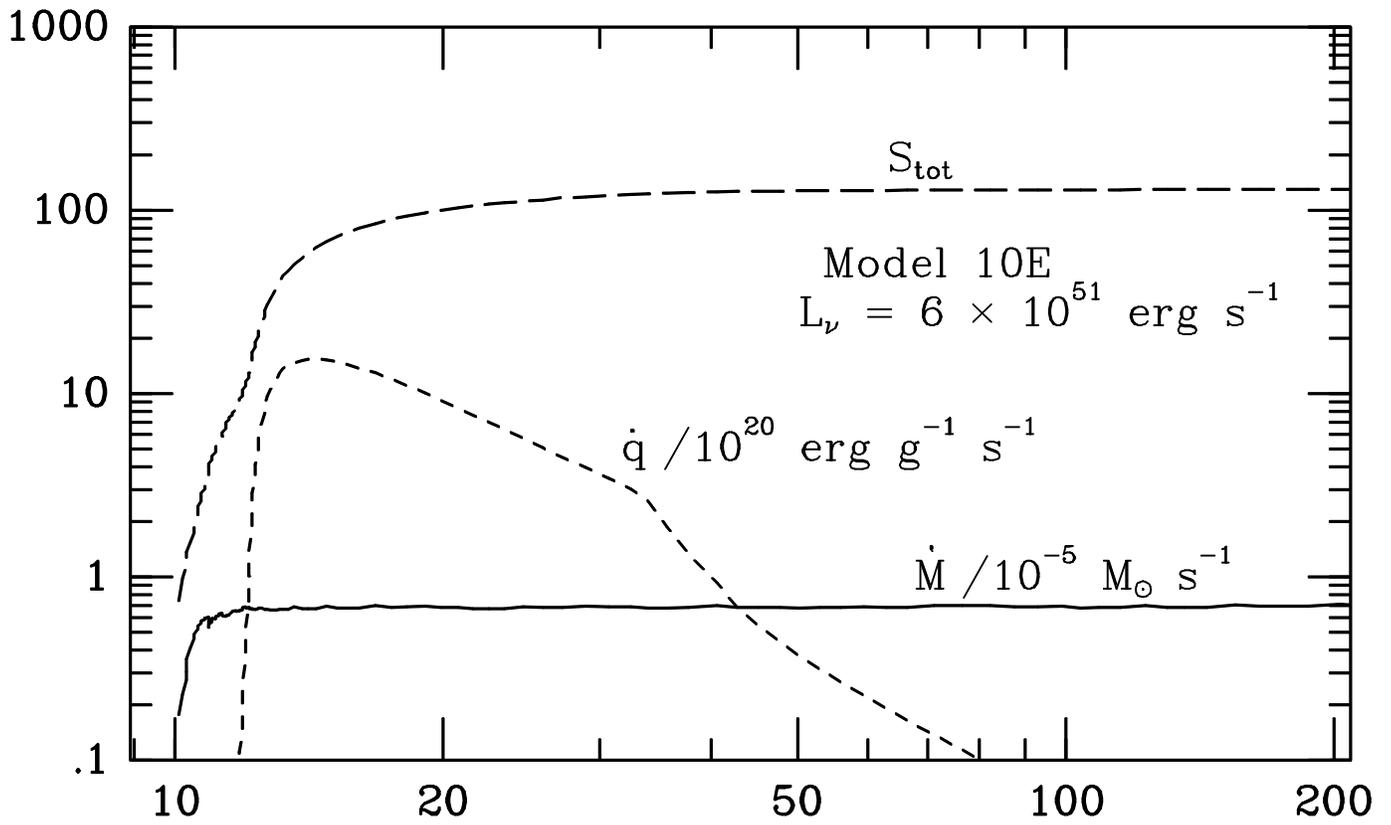
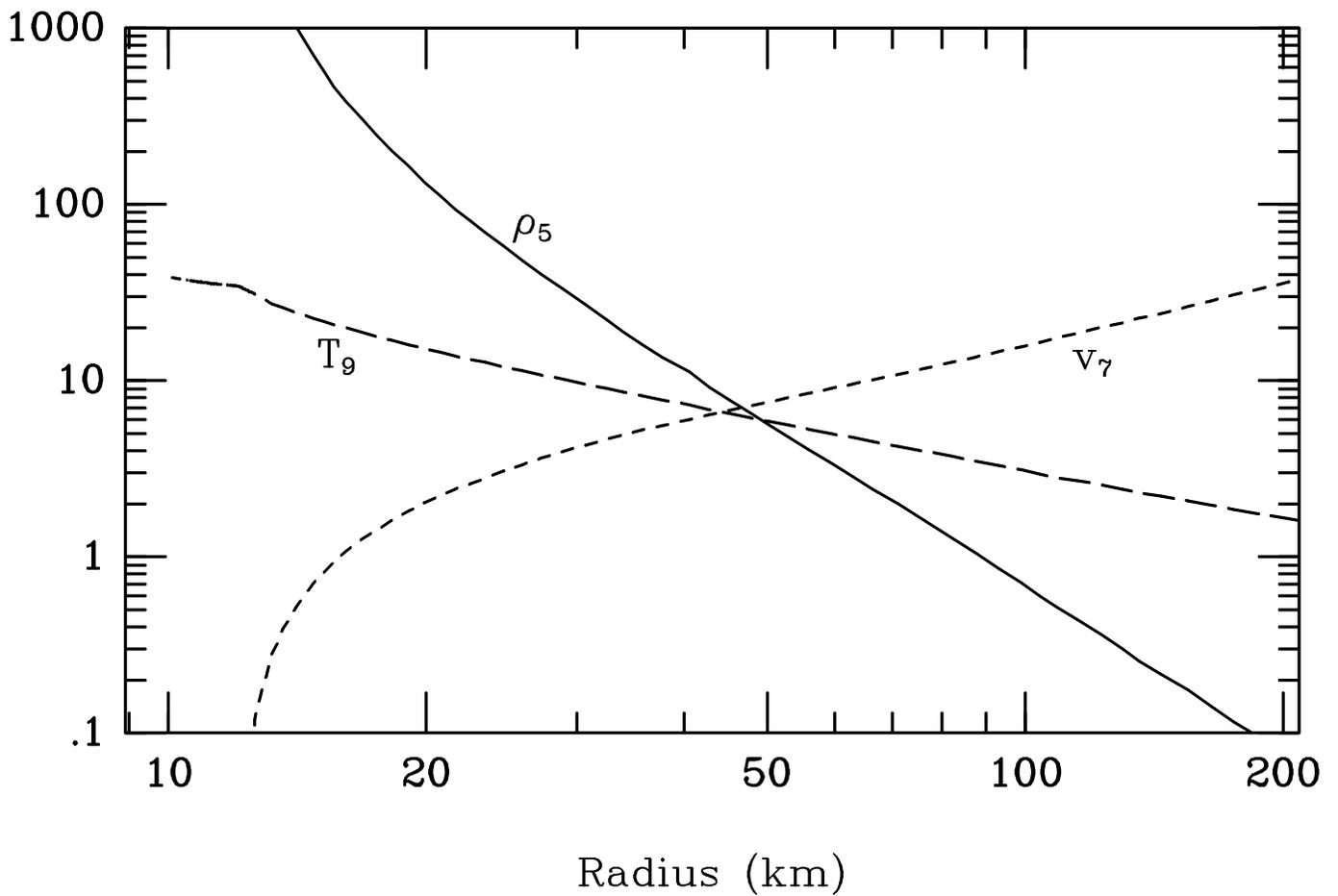
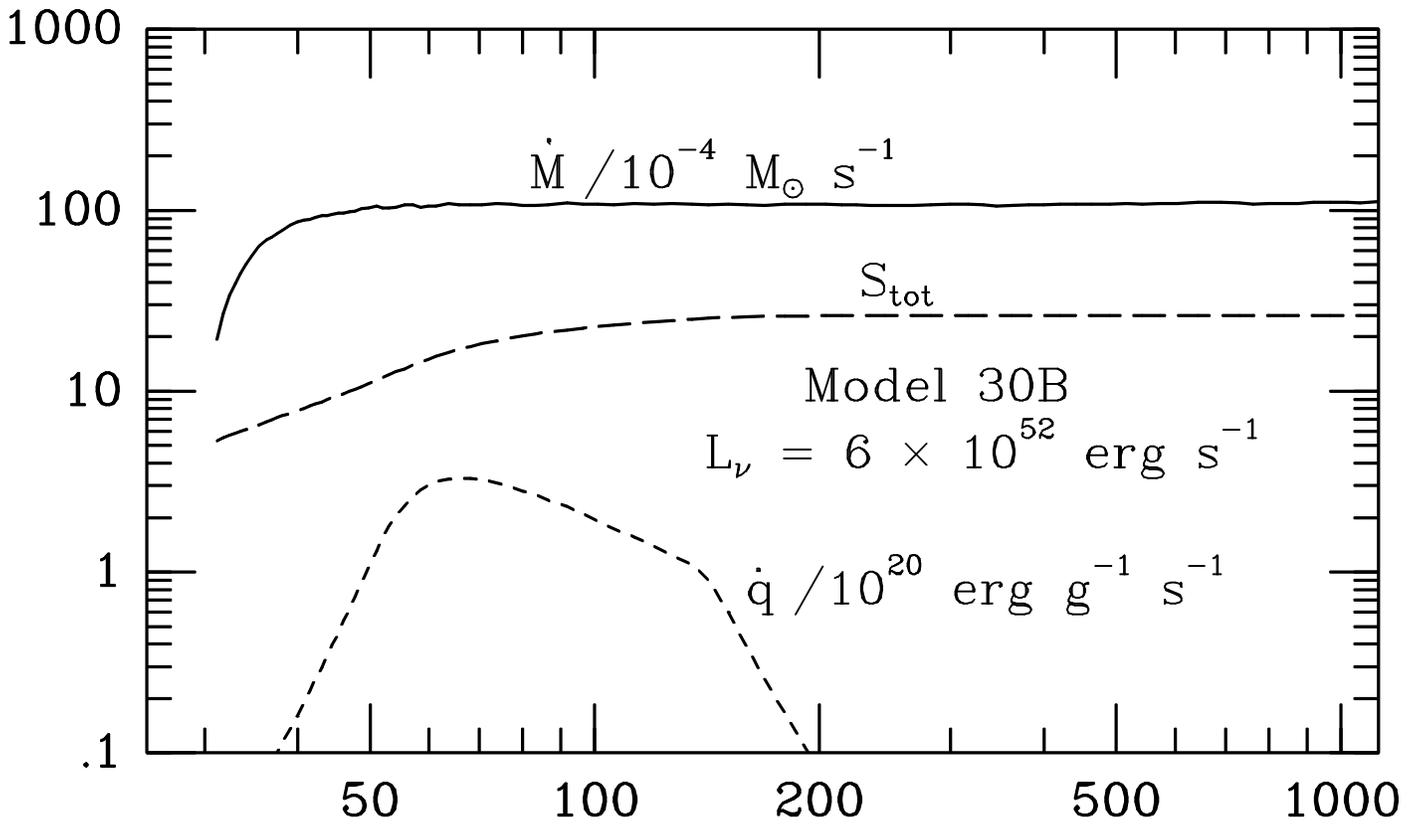
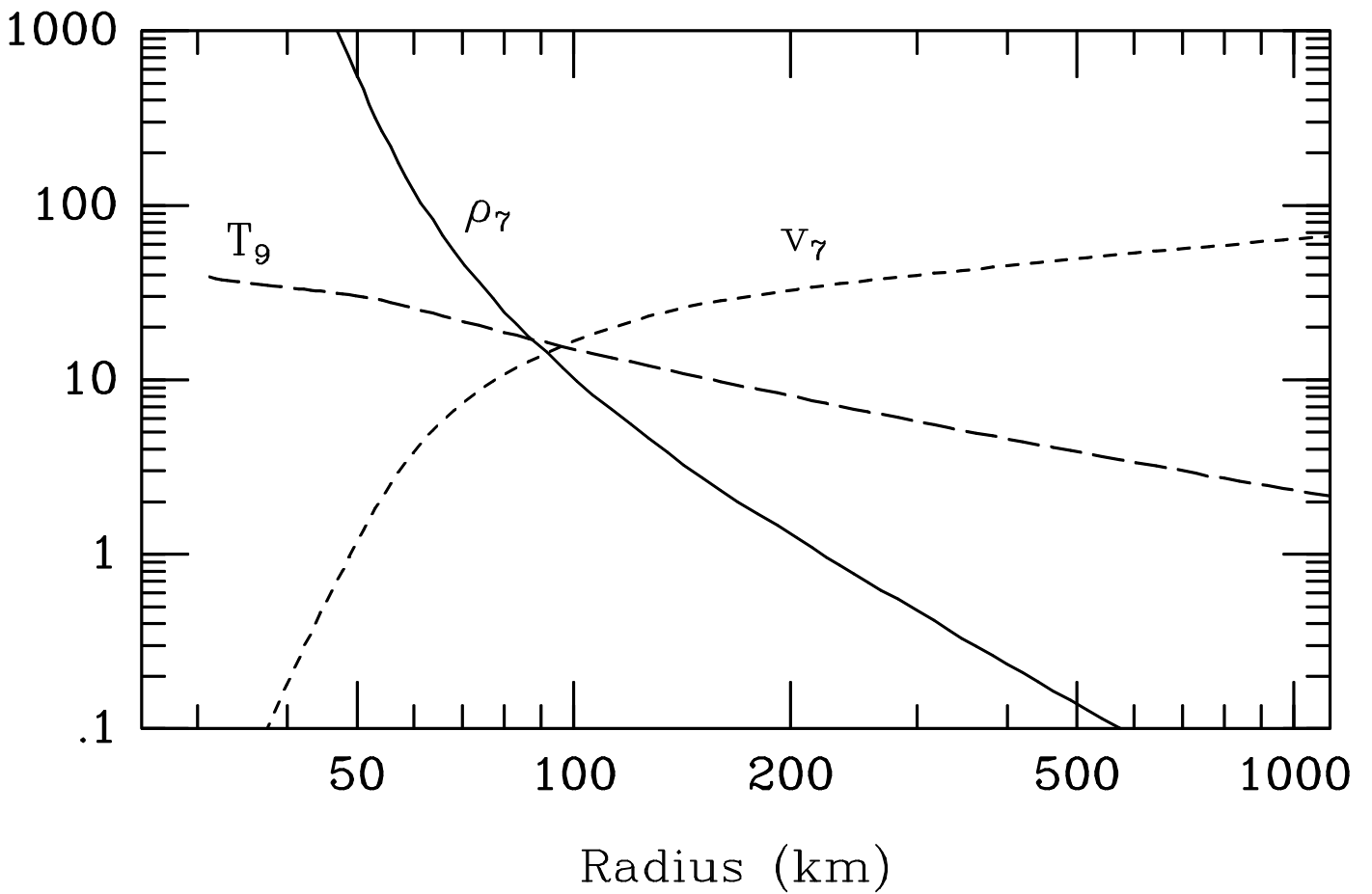
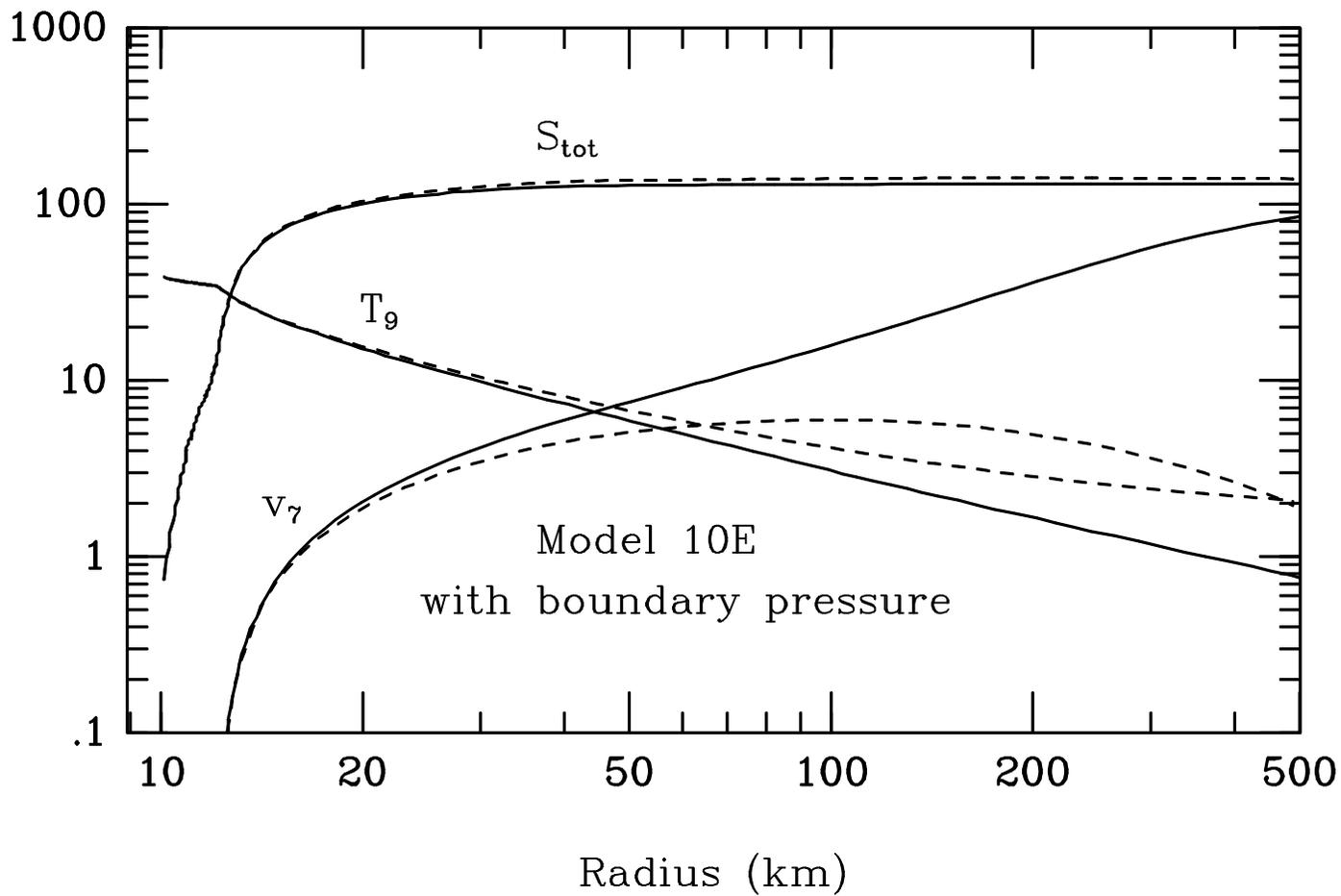
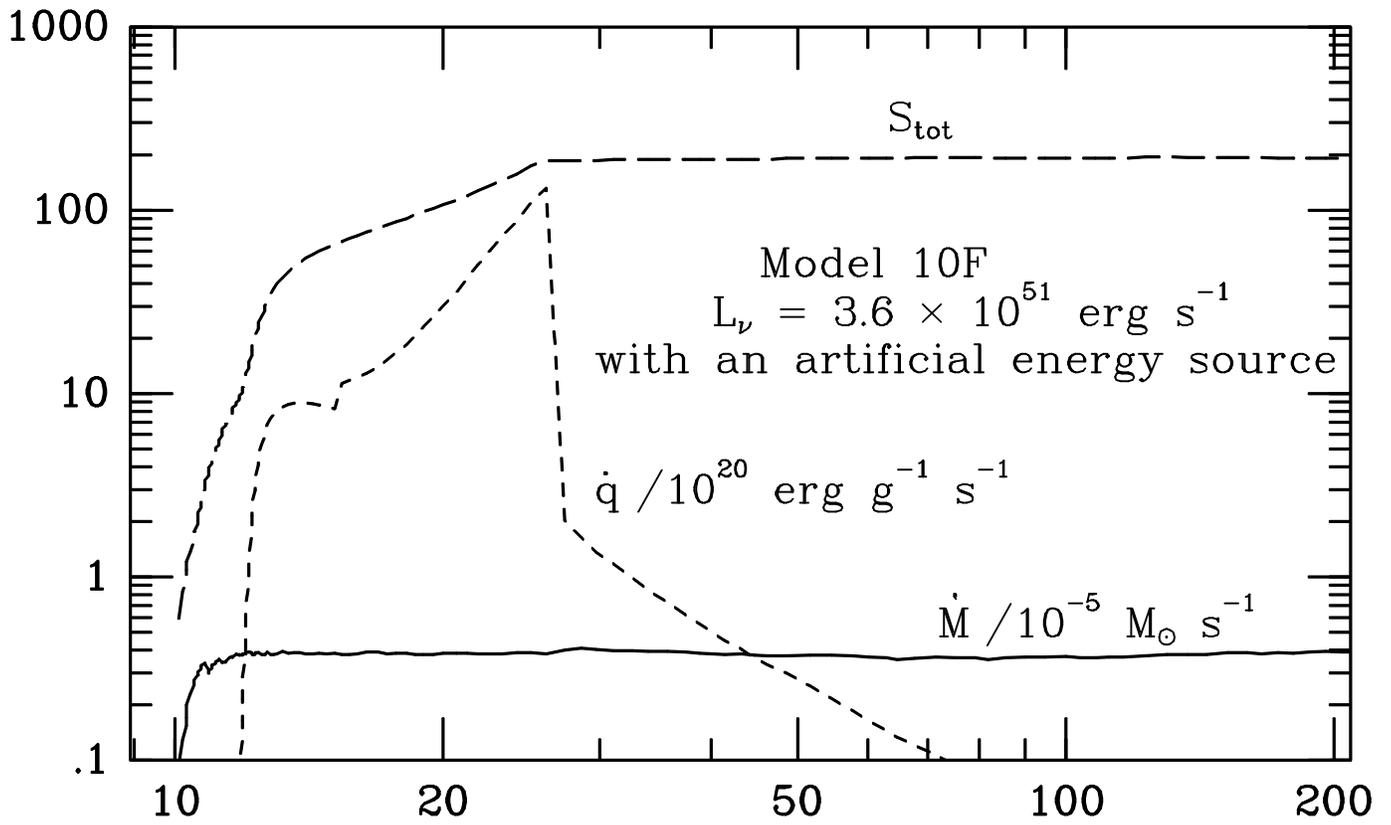
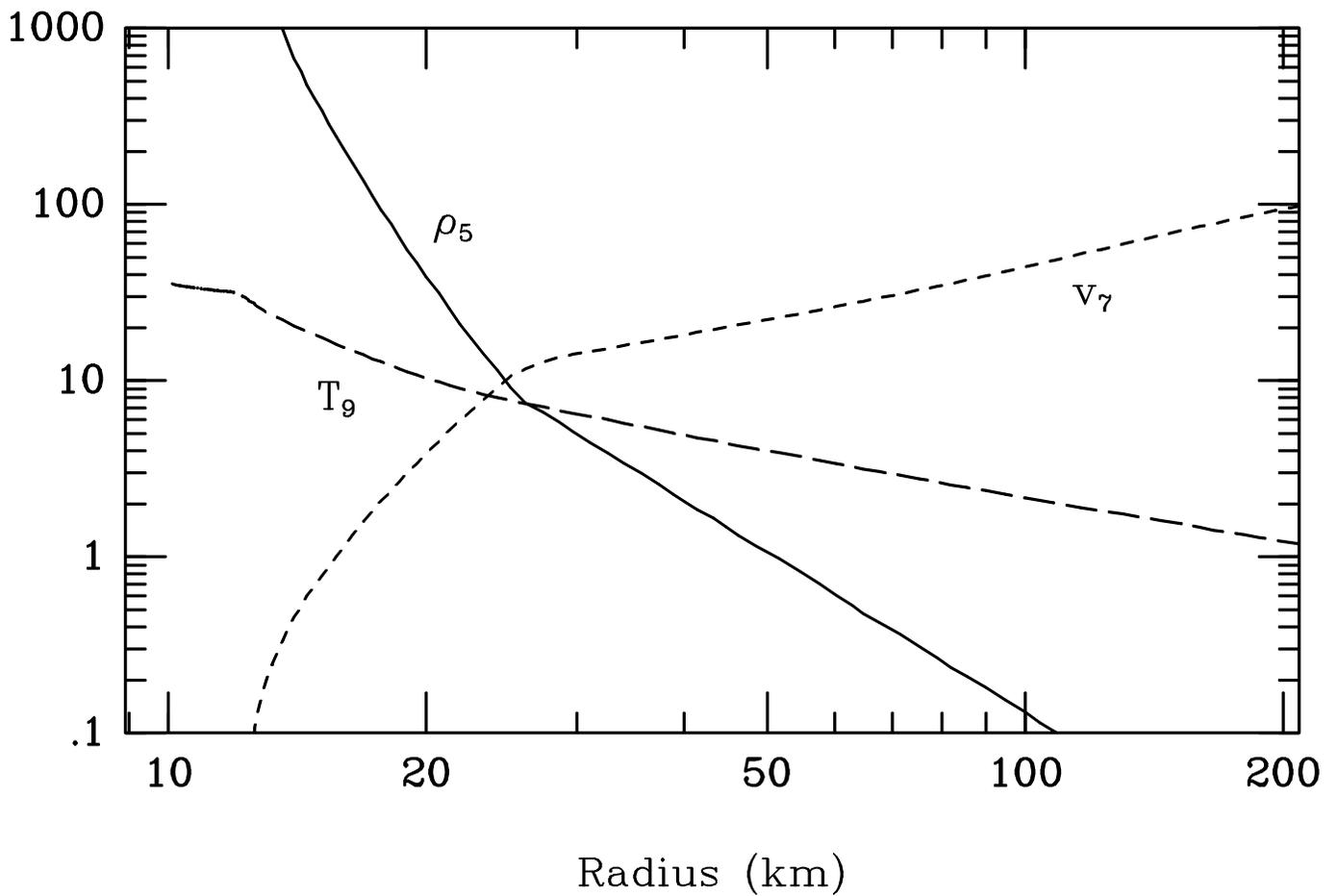
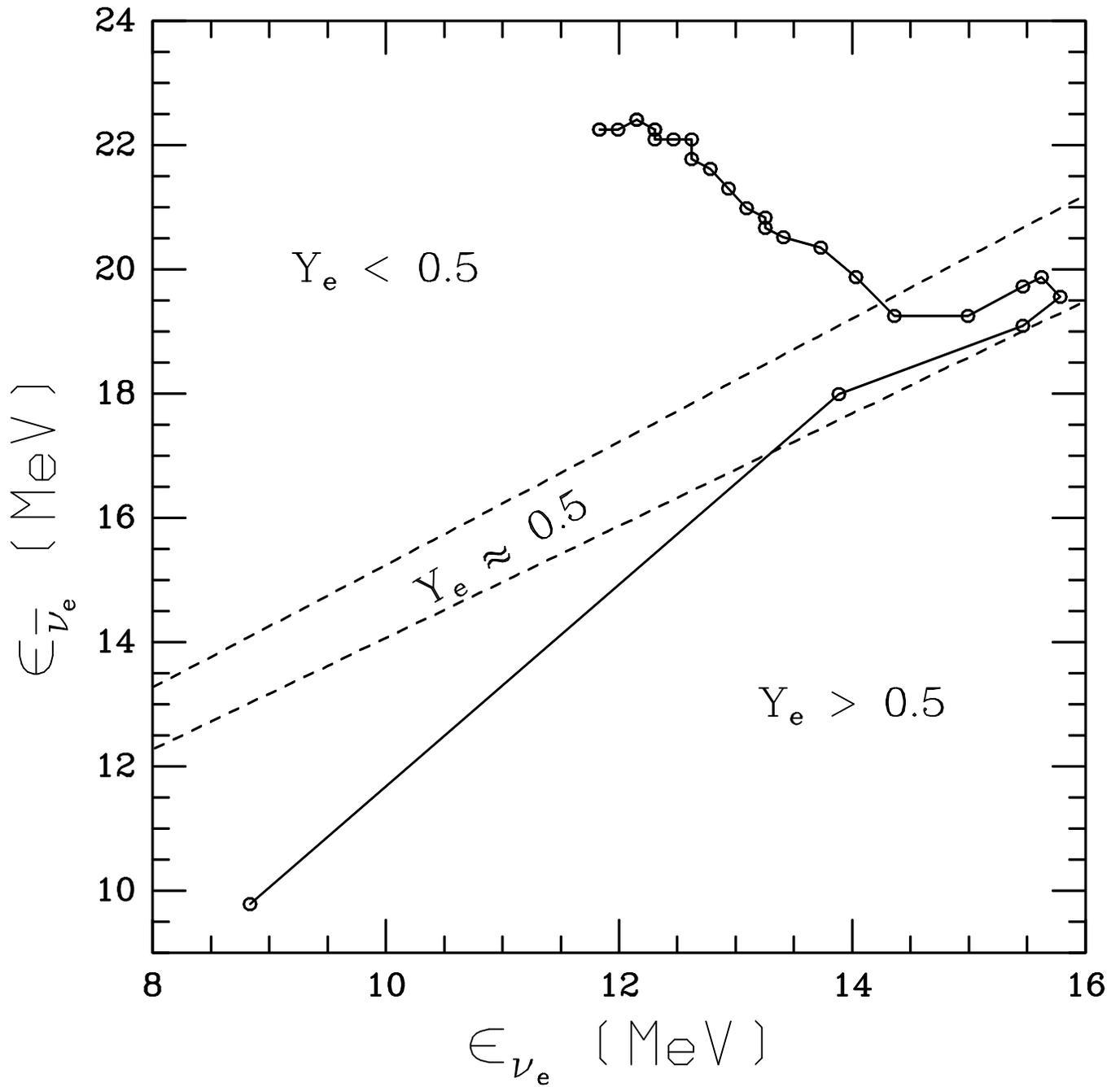
\end{document}